\documentclass[11pt,a4paper]{article}

\usepackage[T1]{fontenc}            
\usepackage[utf8]{inputenc}         
\usepackage{lmodern}                
\usepackage{microtype}              

\usepackage[a4paper, margin=0.7in]{geometry}   

\usepackage{fancyhdr}               
\usepackage{setspace}
\usepackage{titlesec}
\titleformat{\section}{\Large\bfseries\sffamily}{\thesection}{1em}{}
\titleformat{\subsection}{\large\bfseries\sffamily}{\thesubsection}{1em}{}

\usepackage{xcolor}
\usepackage{hyperref}
\hypersetup{
    colorlinks=true,
    linkcolor=blue,                
    citecolor=red,                 
    urlcolor=black                
}
\usepackage[en-US]{datetime2}
\DTMlangsetup*{showdayofmonth=false}
\usepackage[numbers]{natbib}         

\usepackage{amsmath}                
\usepackage{amssymb}                
\usepackage{amsfonts}  
\usepackage{ mathtools}
\usepackage{url}
\usepackage{bm}
\usepackage{float}
\usepackage{textgreek}
\usepackage{physics}
\usepackage{siunitx}
\usepackage[separate-uncertainty=true]{siunitx}
\DeclareUnicodeCharacter{2212}{\textminus}
\DeclareUnicodeCharacter{039B}{$\Lambda$}
\DeclareUnicodeCharacter{2131}{\mathcal{F}}
\DeclareUnicodeCharacter{0301}{\'{e}}
\DeclareUnicodeCharacter{FE20}{-}  
\DeclareUnicodeCharacter{FE21}{-}  
\DeclareUnicodeCharacter{223C}{\sim}
\usepackage{graphicx}               
\usepackage{float}                  
\usepackage{booktabs}               
\usepackage{caption}                
\usepackage{subcaption}             
\usepackage{microtype}
\usepackage{multirow}
\usepackage{pgfplots}
\usepackage{tikz}
\usetikzlibrary{decorations.pathmorphing}
\pgfplotsset{compat=1.18}
\usepackage{listings}               
\usepackage{graphicx}               
\usepackage{svg}                    
\newcommand{\orcidicon}{\includegraphics[width=10pt]{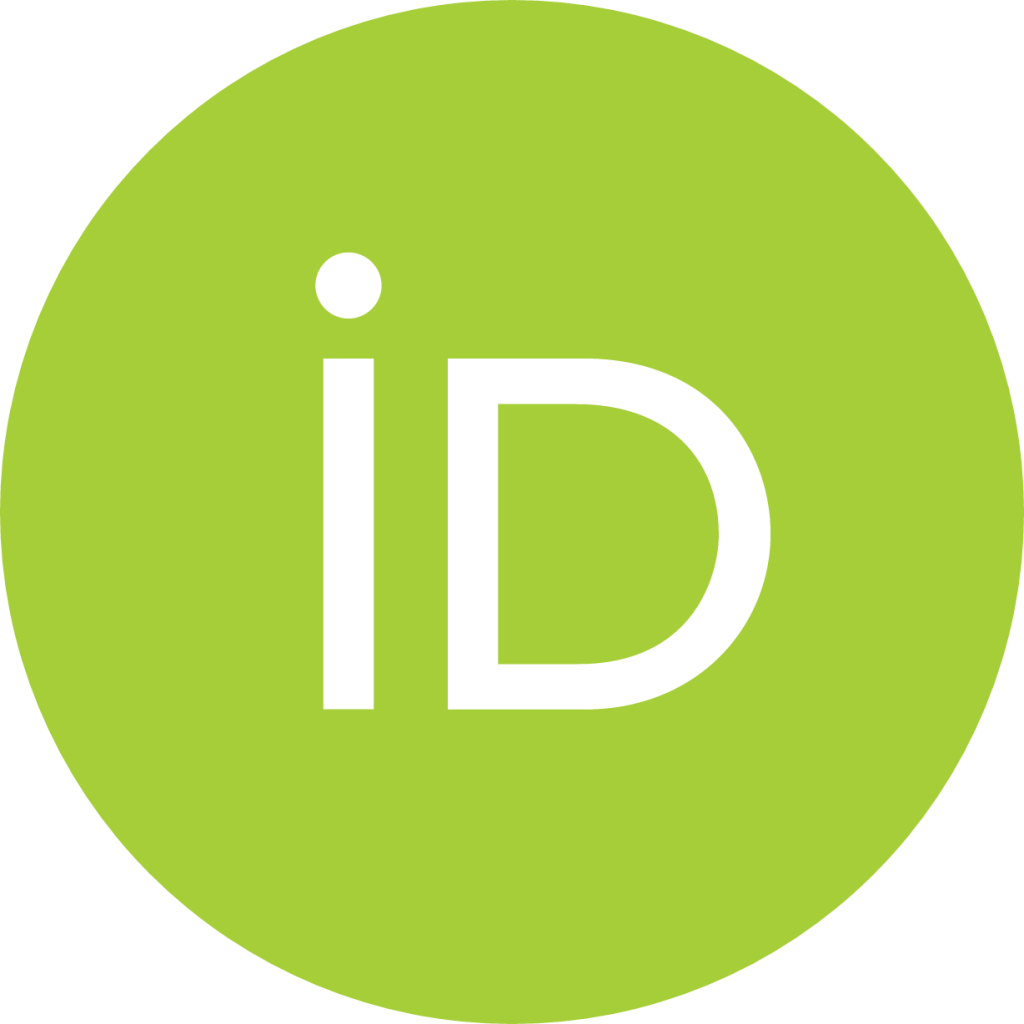}} 

\usepackage{xcolor}
\definecolor{myblue}{RGB}{52, 101, 164}    

\title{\sffamily\bfseries Thermodynamics of Modified Chaplygin-Jacobi Gas and Modified Chaplygin-Abel Gas: Stability Analysis and Observational Constraints}
\author{
    Banadipa Chakraborty\footnote{Department of Physics, Sister Nivedita University, DG Block (Newtown) 1/2, Action Area I, Kolkata-700156, India. Email: \href{mailto:jhelum.chakraborty@gmail.com}{jhelum.chakraborty@gmail.com}}~\href{https://orcid.org/0009-0008-2378-0544}{\orcidicon}, 
    Tamal Mukhopadhyay\footnote{Department of Physics, Sister Nivedita University, DG Block (Newtown) 1/2, Action Area I, Kolkata-700156, India. Email: \href{mailto:tamalmukhopadhyay7@gmail.com}{tamalmukhopadhyay7@gmail.com}}~\href{https://orcid.org/0000-0001-9843-906X}{\orcidicon}, 
    Debojyoti Mondal\footnote{Department of Mathematics, Indian Institute of Engineering Science and Technology, Shibpur, Howrah-711103, India. Email: \href{mailto:djymndl07@gmail.com}{djymndl07@gmail.com}}~\href{https://orcid.org/0000-0002-5370-0778}{\orcidicon}, 
    Ujjal Debnath\footnote{Department of Mathematics, Indian Institute of Engineering Science and Technology, Shibpur, Howrah-711103, India. Email: \href{mailto:ujjaldebnath@gmail.com}{ujjaldebnath@gmail.com}}~\href{https://orcid.org/0000-0002-2124-8908}{\orcidicon}
}
\date{\today}

\begin{document}

\maketitle

\begin{abstract}
This paper explores the thermodynamic properties and stability of two newly introduced gas models, namely the Modified Chaplygin-Jacobi gas and the Modified Chaplygin-Abel gas. To achieve this, we examine the behavior of relevant physical parameters to gain in depth information about the evolution of the universe. The specific heat formalism is employed to verify the applicability of the third law of thermodynamics. Furthermore, the equation of state for the thermal system is obtained by applying thermodynamic variables. The stability of the gas models is investigated within the framework of classical thermodynamics, focusing on adiabatic processes, specific heat capacities, and isothermal conditions. It is inferred that the proposed fluid configurations exhibit thermodynamic stability and undergo adiabatic expansion for suitable parameter choices. We then perform observational analysis using CC+BAO and Pantheon+SH0ES datasets to impose constraints on our model parameters using the Markov Chain Monte Carlo (MCMC) process.
\end{abstract}

\section{Introduction}\label{Sect:Intro}
The present cosmological observations from the Type Ia supernovae indicate our universe to be expanding in an accelerated manner \cite{riess1998observational, Perlmutter_1998, perlmutter1999measurements} and hence contradict the strong energy condition, which implies $\rho+3p < 0$. The exotic matter that drives the universe to expand acceleratedly is known to be ``Dark Energy'' \cite{peebles2003cosmological,padmanabhan2003cosmological}. The straightforward approach to resolve this problem was to take the famous cosmological constant $\Lambda$ as a suitable choice for dark energy taking the equation of state (EoS) parameter $\omega=\frac{p}{\rho}=-1$ as constant. However this approach has faced several fatal drawbacks that remain unanswered \cite{velten2014aspects,yoo2012theoretical,weinberg2001cosmological}.  To overcome this problem, several dark energy models with dynamic vacuum energy have been proposed.  This theory introduces an additional degree of freedom, characterized by a scalar field. The Lagrangian incorporates a potential or kinetic term, which facilitates the transition of the universe from a dust-like matter-dominated state to one experiencing accelerated expansion. The quintessence model \cite{peebles1988cosmology, Caldwell_1998,zlatev1999quintessence,carroll1998quintessence} is one that kind that is mostly favored by physicists to be a potential candidate for dark energy. Additionally, alternative dark energy models featuring dynamic vacuum energy have been explored, including tachyon fields \cite{Sen_2002}, k-essence \cite{armendariz2001essentials}, H-essence \cite{Wei_2005}, dilaton \cite{Gasperini_2001}, phantom fields \cite{Caldwell_2002}, DBI-quintessence \cite{Gumjudpai_2009}, and DBI-essence \cite{martin2008dbi}. The Chaplygin gas model, initially introduced in the context of aerodynamics by Sergey A. Chaplygin in 1904~\cite{chaplygin1944gas}, has recently attracted considerable interest in both cosmology and theoretical physics because of its distinctive equation of state (EoS). The consideration of using Chaplygin gas in a FRW universe for the explanation of the accelerated expansion of the universe was first proposed by Kamenshchik et al. \cite{kamenshchik2001alternative,gorini2005chaplygin}. Chaplygin gas equation of state reveals a fascinating link to string theory, as it originates from the Nambu-Goto action, which governs the dynamics of D-branes moving within a $(d + 2)$-dimensional spacetime under the light-cone gauge~ \cite{bordemann1993dynamics}. In addition, the Chaplygin gas is the only known fluid that allows for a supersymmetric extension~ \cite{hoppe1993supermembranes, jackiw2000supersymmetric}. The Chaplygin gas-like equation of state can then be derived from a tachyon scalar field model, which is described by a Dirac-Born-Infeld type Lagrangian with a constant potential for the field~\cite{frolov2002prospects,gorini2004tachyons}. A significant aspect of the Chaplygin gas model is its ability to unify dark matter and dark energy within a single theoretical framework~\cite{fabris2008matter,bento2004revival}. But, this model needed to explain the structure formation era in the universe \cite{bean2003chaplygin,sandvik2004end}. Bento et al. \cite{bento2002generalized} introduced a more general version of the Chaplygin gas EoS and it is referred to as Generalised Chaplygin gas (GCG). Another modification to the equation of state was introduced and is known as Modified Chaplygin Gas(MCG)\cite{benaoum2022accelerated, debnath2004role}. The MCG model has limitations in accurately describing the universe's behavior between the early and late acceleration periods \cite{PhysRevD.92.103511}. To improve the agreement between models and observational data, various limits on the model parameters of both the Chaplygin Gas (CG) and Modified Chaplygin Gas (MCG) were explored \cite{ranjit2013observational,paul2013observational, bhadra2014constraining,paul2016constraints,panigrahi2015constraining,mebarki2019dynamical}. Villanueva proposed the generalized Chaplygin-Jacobi gas through the application of Jacobi's elliptic function~ \cite{villanueva2015generalized,villanueva2015jacobian}, which is linked to the generalized Chaplygin scalar field. This model is capable of describing the inflationary phase and has been tested against the data recently provided by the Planck 2015 mission. Recently, Debnath \cite{debnath_MCAG_MCJG} introduced two models of Chaplygin gas namely, Modified Chaplygin-Jacobi Gas (MCJG) and Modified Chaplygin-Abel Gas (MCAG). He replaced the hyperbolic cosine function in the generating function of the scalar field with the Jacobi and Abel elliptic functions, deriving the equation that connects the pressure and density of the MCJG and MCAG. The reason behind using such elliptic functions is as follows: The exact solution of the Einstein field equations with an inflation field is very much hard to achieve. Still using slow roll approximations and other methods several literature suggests that the inflationary potentials has the form of trigonometric and hyperbolic functions~\cite{PhysRevD.50.4794,kim2013inflationattractorscalarcosmology, chervon1996exact, Campo_2012,Campo_2013,harko2014arbitrary}. The elliptic functions are more generalised functions which reduce to trigonometric and hyperbolic forms as a special case. So, the use of elliptic functions in the inflationary potential and then the further extension of this function to obtain Modified Chaplygin gas like equation of state provide more generalisation, from both mathematical and physical perspectives.
\\The intersection of black hole thermodynamics and Einstein's field equations emerged as a fundamental concept within Einstein gravity framework \cite{jacobson1995thermodynamics}. This connection established that the entropy-area proportionality principle, coupled with the fundamental thermodynamic relation $\delta Q = T dS$ in Rindler spacetime, leads naturally to Einstein's equations. The correspondence between thermodynamic and geometric quantities in black hole physics reveals itself through entropy's relationship to horizon area and temperature's connection to surface gravity \cite{bekenstein1973black}. A significant advancement occurred when Verlinde demonstrated \cite{verlinde2000holographic} that the Friedmann equation in a radiation-dominated FRW universe exhibits structural similarities to the Cardy-Verlinde formula, which describes entropy relationships in conformal field theory. The thermodynamic properties of de Sitter spacetime were first elucidated through the pioneering work of Gibbons and Hawking \cite{gibbons1977cosmological}. Further developments by Gong and colleagues \cite{gong2007j} established the apparent horizon as the most relevant boundary for thermodynamic analysis, based on the established relationship between Friedmann equations and thermodynamic principles. This becomes particularly significant in accelerating universes dominated by dark energy, where the event horizon differs from the apparent horizon. Research has shown that thermodynamic laws maintain validity at the apparent horizon but fail at the event horizon when these boundaries are distinct \cite{wang2006thermodynamics}. Recently, another work has showed that the thermodynamics of any matter fields including the Chaplygin gas is irreversible in nature, when satisfying the Landauer principle~\cite{odintsov2024landauerprinciplecosmologylink}. Investigations into Chaplygin gas-dominated universes \cite{izquierdo2006dark} examined the generalized second law (GSL) in relation to event horizons in accelerating cosmological models, confirming GSL validity during early expansion phases. A comprehensive comparative study \cite{chakraborty2019thermodynamics} evaluated GSL validity at both cosmological horizons (apparent and event) in FRW universes containing various Chaplygin gas models. The thermodynamic framework has been extensively explored across multiple variations:
\begin{itemize}
\item Standard Chaplygin gas \cite{myung2011thermodynamics}
\item Generalized Chaplygin gas \cite{santos2006thermodynamic}
\item Modified Chaplygin gas \cite{debnath2004role, bandyopadhyay2010laws, santos2007thermodynamic}
\item Variable modified Chaplygin gas \cite{chakraborty2019evolution}
\end{itemize}
This extensive body of research provides the foundation for our investigation into the thermodynamic characteristics of MCJG and MCAG parameters.
\\The paper aims to analyse the thermodynamic behaviour of MCJG and MCAG and also study their temperature behaviour. Also the thermodynamic stability of MCJG and MCAG using thermal and adiabatic equation of states is part of our analysis in this work. The paper is arranged as follows: In Sect.~\ref{Sect:MCJG} and \ref{Sect:MCAG} we have approximate MCJG and MCAG model and analysed their thermodynamic behaviour and stability, Sect.~\ref{Sect:Observational Study} is dedicated to study the model parameters and testing the viability of our approximations using CC+BAO and Pantheon+SH0ES datasets and finally we wind up our paper in Sect.~\ref{Sect:conclusion} with concluding remarks.

\section{Thermodynamic Analysis of Modified Chaplygin-Jacobi Gas}\label{Sect:MCJG}

\begin{figure}[htbp]
    \centering
    \begin{subfigure}[b]{0.45\textwidth}
        \includegraphics[width=1.03\textwidth]{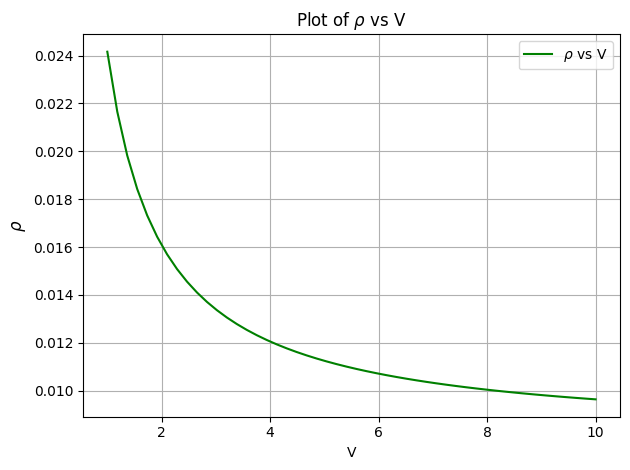}
        \caption{}
        \label{fig:rho_MCJG}
    \end{subfigure}
    \begin{subfigure}[b]{0.45\textwidth}
        \includegraphics[width=1.1\textwidth]{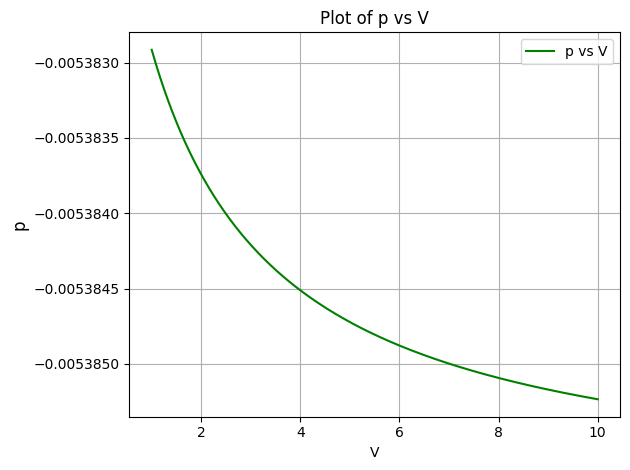}
        \caption{}
        \label{fig:p_MCJG}
    \end{subfigure}
    \caption{A plot showing the dark energy density $\rho$ and pressure $p$ as functions of volume $V$ for the Modified Chaplygin-Jacobi Gas model.}
    \label{fig:rho_p_MCJG}
\end{figure}

Jacobi elliptic functions represent a crucial category of elliptic functions. By substituting the hyperbolic cosine function in the Hubble parameter with a Jacobi elliptic function, the relationship between pressure and density can be expressed as~\cite{debnath_MCAG_MCJG},

\begin{equation}\label{MCJG_EoS}
p = \left[(2K - 1)(1 + A) - 1\right]\rho - \frac{KB}{\rho^\alpha} + \frac{(1 - K)(1 + A)^2 \rho^{2 + \alpha}}{B} \quad .             
\end{equation}
In this context, $A$ and $B$ are positive constant parameters, with $0 \leq \alpha \leq 1$ and $0 \leq K \leq 1$ representing the modulus of the elliptic function. This expression defines the pressure-density relationship of the ``Modified Chaplygin-Jacobi Gas''. The constant parameters $A$, $B$ and $\alpha$ are coming from the Modified Chaplygin gas and generalised Chaplygin gas equation of state~\cite{debnath2004role,bento2002generalized} and the detailed observational constraints of these parameters are given in ref~\cite{Benoum_mcg,Amendola_2003, Lu_2010,Zhu_gcg,lu2009observational}. Here, $B$ is a positive constant that influences the pressure term inversely proportional to the energy density. It plays a significant role in the thermodynamic stability and the transition between different cosmological eras~\cite{Thakur_2009,adhav2011statefinder}\footnote{In the aforementioned references this parameter is identified as $A$, but physically that does not make any difference as long as the form of the EoS of the MCG and GCG is preserved. Readers can tally the form of the EoS of MCG with ref.~\cite{debnath2004role} to avoid any kind of confusion.}. The value of $B$ affects the stability of the model and the collapse time rate in structure formation scenarios~\cite{karbasi2015}. $A$ is a positive constant that modifies the linear term in the equation of state. It is crucial for describing the current accelerated expansion of the universe and is constrained by observational data such as CMBR measurements and The range of $A$ is typically constrained to ensure compatibility with observational data, with values often found between $-0.35$ and $0.025$~\cite{Thakur_2009,Liu_Dao_Jun_2005}\footnote{Here, also the same goes for the parameter $A$. The authors of the above references use $B$ instead of $A$ but physically they both signify the same thing.}. For the values of $A=\frac{1}{3}$ the MCG model can reproduce radiation dominated era for minuscule values of scale factor. In some models, $A$ can be a function of the scale factor, adding variability to the equation of state and affecting the stability and thermodynamic properties of the gas~\cite{Panigrahi_2016}. The $\alpha$ parameter determines the non-linearity of the inverse density term. It ranges between $0$ and $1$ and influences the transition from radiation-dominated to quintessence-dominated eras~\cite{adhav2011statefinder,adhav2011statefinder2}. The value of $\alpha$ affects the thermodynamic stability and the sound speed of the gas, with higher values leading to different stability conditions~\cite{Ferreira2018,santos2007thermodynamic} and it also plays a role in the evolution of perturbations and the turn-around redshifts in structure formation models~\cite{adhav2011statefinder}. The choice of the elliptic modulus $K=1$ reduced the  generating function of MCJG to the generating function of MCG~\cite{debnath_MCAG_MCJG}.
From the general Thermodynamics we know that\cite{landau1984statistical},
\begin{equation}\label{rho_basic_definition}
\rho = \frac{U}{V} \quad,               
\end{equation}
and,
\begin{equation}\label{p_basic_definition}
 p = -\left(\frac{\partial U}{\partial V}\right)_S .               
 \end{equation}
From \eqref{MCJG_EoS} and \eqref{p_basic_definition} we get the expression for the internal energy U as,

\begin{equation}\label{internal energy_MCAG}
U=V \left( \dfrac{B \left(1 + \dfrac{1}{-1 + K - e^{(1 + A)(1 + \alpha) B C_1} V^{(A+1)(\alpha+1)}}\right)}{1 + A} \right)^{\frac{1}{1 + \alpha}}.
\end{equation}
Here, $C_1$ is an integration constant and $C_1=C_1(S)$. Now If we put $A=0$, $K=1$ in the eq.~\eqref{MCJG_EoS} then we get,
\[p = -\dfrac{B}{\rho^{\alpha}} \quad,\]
which is nothing but the equation of state for the Generalized Chaplygin Gas~\cite{bento2002generalized}. Now comparing the energy density of GCG with that of MCJG we have, 
\begin{equation}\label{MCJG_GCG_equivalence_condition}
    e^{-BC_1(1+\alpha)}=-\dfrac{C}{B} \quad.
\end{equation}
Here, $C$ is an arbitrary parameter which is a function of entropy $(S)$ only. Now putting this relation into eq.~\eqref{internal energy_MCAG} and using eq.~\eqref{rho_basic_definition} we have the energy density in the form

\begin{equation}
\rho =  \left[ \frac{B}{1 + A} \left(1 - \dfrac{1}{1 - K - (\frac{B}{C})^{(A+1)} V^{(A+1)(\alpha+1)}}\right) \right]^{\frac{1}{1 + \alpha}} .
\end{equation}                    
Now, by further simplifications we will arrive at, 
\begin{equation}\label{rho_MCJG}
    \rho = \left[ \frac{B}{A+1} \left(1 + 2K\left(\dfrac{V_0}{V}\right)^{(\alpha+1)(A+1)}\right) \right]^{\dfrac{1}{1 + \alpha}} .
\end{equation}
Here, we have used 
\begin{equation}\label{V0_MCJG}
    V_0^{(\alpha+1)} = \dfrac{C}{B} \quad.
\end{equation}
Now,  From equation \eqref{rho_MCJG}, it is clearly obtainable that at a very low volume, i.e., for $V  \to 0$, the energy density becomes very high. Conversely, in the limit of very large volume, i.e., as $V \to \infty$, the energy can be interpreted as a combination of two fluids: one exhibiting a constant energy density and the other exhibiting a volume-dependent energy density, expressed as follows:
\begin{equation}\label{rho_MCJG_high_volume}
   \rho \approx \left[\left(\dfrac{B}{A+1}\right)^{\frac{1}{1+\alpha}} + \dfrac{2K}{\alpha+1}\left(\dfrac{V_0}{V}\right)^{(A+1)(\alpha+1)}\right] .
\end{equation}
Now, the pressure of this MCJG gas can be obtained by,
\begin{equation}\label{p_MCJG}
   p = -B K \left[ \frac{B}{A + 1} \left( 1 + \dfrac{\left( \dfrac{V_0}{V} \right)^{(A + 1)(\alpha + 1)}}{1 + K \left( \dfrac{V_0}{V} \right)^{(A + 1)(\alpha + 1)}} \right) \right]^{-\dfrac{\alpha}{\alpha + 1}}. 
\end{equation}
From this equation, we can write that for Small Volume (and High Energy Density) pressure becomes zero which means the gas behaves as pressure-less. 
\begin{align}\label{p_MCJG_small_volume}
    p &\approx -BK \left[\left(\dfrac{B}{A+1}\right)^{\frac{1}{\alpha + 1}} + \left(1+\dfrac{1}{K(\alpha +1)}\right)\right]^{-\alpha} \nonumber\\
    & \approx -K \left(\dfrac{B}{A+1}\right)^{\frac{1}{\alpha +1}} + K \left(\dfrac{B}{A+1}\right)^{\frac{1}{\alpha +1}} \nonumber\\
    & \approx 0,
\end{align}
and, at Large Volume (i.e. at Low Energy Density), the pressure becomes a negative constant which resembles a cosmological constant scenario.
\begin{equation}\label{p_MCJG_high_volume}
    p \approx -K \left(\dfrac{B}{A+1}\right)^{\frac{1}{\alpha +1}}. 
\end{equation}
The similar behaviour of energy density and pressure is depicted in fig.\ref{fig:rho_MCJG} and \ref{fig:p_MCJG}.

\subsection{EoS Parameter}
The equation of state parameter (EoS) parameter is defined by $$ \omega=\dfrac{p}{\rho}\quad.$$ From the equations \eqref{p_MCJG} and \eqref{rho_MCJG} we have obtained, 
\begin{equation}\label{EoS_MCJG}
    \omega = - \dfrac{B K}{\left[ \dfrac{B }{A + 1}\left( 2K \left( \dfrac{V_0}{V} \right)^{(A + 1)(\alpha + 1)} + 1 \right) \right]^{\frac{1}{\alpha + 1}} \left[ \dfrac{B \left( K \left( \dfrac{V_0}{V} \right)^{(A + 1)(\alpha + 1)} + \left( \dfrac{V_0}{V} \right)^{(A + 1)(\alpha + 1)} + 1 \right)}{(A + 1) \left( K \left( \dfrac{V_0}{V} \right)^{(A + 1)(\alpha + 1)} + 1 \right)} \right]^{\frac{\alpha}{\alpha + 1}}}\quad.
\end{equation}
It represents a very small constant value of EoS parameter at small volume ($V \to 0$) and at large volume ($V \to \infty$) the EoS parameter becomes negative and volume dependent and is plotted in fig.~\ref{fig:omega_MCJG}. It is interesting to show that for the value of parameters $A=V_0=0$, $B=K=1$ the EoS expressed in \eqref{EoS_MCJG} reduces to $\omega=-1$ which corresponds to the original dark energy equation of state according to the $\Lambda$CDM model for any values of $\alpha$.

\subsection{Deceleration Parameter}
\begin{figure}[htbp]
    \centering
    \begin{subfigure}[b]{0.45\textwidth}
        \includegraphics[width=1.03\textwidth]{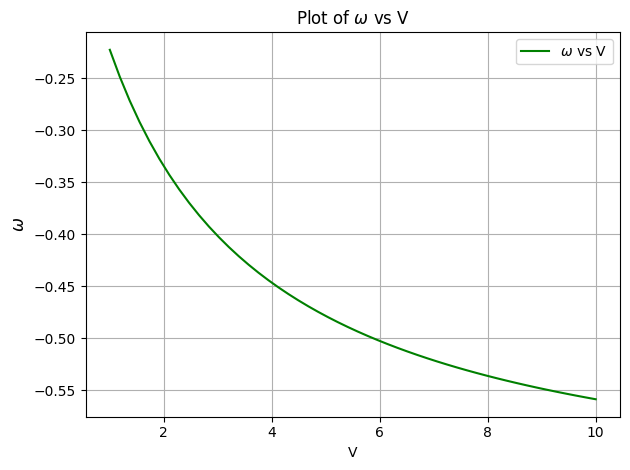}
        \caption{}
        \label{fig:omega_MCJG}
    \end{subfigure}
    \begin{subfigure}[b]{0.45\textwidth}
        \includegraphics[width=1.05\textwidth]{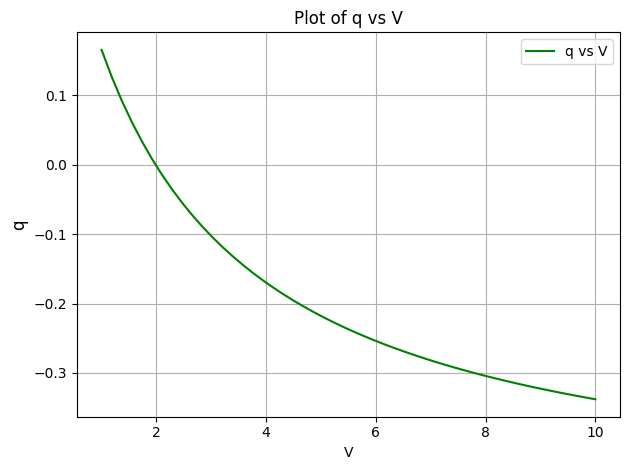}
        \caption{}
        \label{fig:q_MCJG}
    \end{subfigure}
    \caption{A graph depicting the equation of state parameter $\omega$ and the deceleration parameter $q$ as functions of volume $V$ for the Modified Chaplygin-Jacobi Gas model.}
    \label{fig:omega_q_MCJG}
\end{figure}

The deceleration parameter $q$ characterizes the rate at which the universe expands. A negative value of $q$ signifies that the universe is experiencing accelerated expansion, whereas a positive value indicates a phase of decelerated expansion. The deceleration parameter is given by:

\begin{equation}\label{q_definition}
q = \frac{1}{2} + \frac{3p}{2\rho}\quad.
\end{equation}
Putting the values of $p$ and $\rho$, the $q$ parameter for MCJG takes the form,
\begin{equation}\label{q_MCJG}
    q = - \dfrac{3 B K}{2 \left[ \frac{B }{A + 1} \left( 2K \left( \dfrac{V_0}{V} \right)^{(A + 1)(\alpha + 1)} + 1 \right)\right]^{\frac{1}{\alpha + 1}} \left[ \frac{B }{A + 1}\left( \dfrac{\left( \dfrac{V_0}{V} \right)^{(A + 1)(\alpha + 1)}}{K \left( \dfrac{V_0}{V} \right)^{(A + 1)(\alpha + 1)} + 1} + 1 \right) \right]^{\frac{\alpha}{\alpha + 1}}} + 0.5\quad .
\end{equation}
Now, by considering the limiting case:
\begin{align}\label{q_MCJG_limiting}
\lim_{V \to 0}q &= 0.5, \\
\lim_{V \to \infty}q &= -\frac{3AK}{2} - \frac{3K}{2} + \frac{1}{2}\quad.
\end{align}
Here, when the volume is small, then $\frac{V}{V_0}$ term becomes small and equals to $0.5$, making the deceleration parameter positive and for large volume  $\frac{V}{V_0}$ term dominates and becomes larger than $0.5$ making the deceleration parameter negative.
Therefore, negative values of the deceleration parameter indicate that the universe is currently experiencing an accelerated expansion at large scales. This transition is visible in the fig.~\ref{fig:q_MCJG}.

\subsection{Square Speed of Sound}\label{subsect:soundsquare}

Here, we examine the classical stability of MCJG against perturbation by using the speed of sound as
\[v_s^2 = \left(\frac{\partial p}{\partial \rho}\right)_S.\] 
Thus, we have, 

\begin{align}\label{v_s^2_MCJG}
v_s^2 = \frac{1}{B} \left(\frac{BY}{A+1}\right)^{\frac{1}{\alpha+1}} 
\left[B^2K\alpha + B((A+1)(2K-1)-1)\left(\frac{BY}{A+1}\right)^{\frac{\alpha+1}{\alpha+1}} \right. \\ \nonumber
\left. - (A+1)^2(K-1)(\alpha+2)\left(\frac{BY}{A+1}\right)^{\frac{2\alpha+2}{\alpha+1}}\right]
\left(\frac{BY}{A+1}\right)^{-\frac{\alpha+2}{\alpha+1}}\quad ,
\end{align}
\noindent where $Y$ is defined as,
\begin{equation}
    Y = 2K\left(\frac{V_0}{V}\right)^{(A+1)(\alpha+1)} + 1 \quad.
\end{equation}
The figure.~\ref{fig:v^2_MCJG} shows that $v_s^2$ is positive throughout the expansion of the universe and hence classically stable under perturbation.

\subsection{Thermodynamic Stability}
\begin{figure}[htbp]
    \centering
    \begin{subfigure}[b]{0.45\textwidth}
        \includegraphics[width=1.03\textwidth]{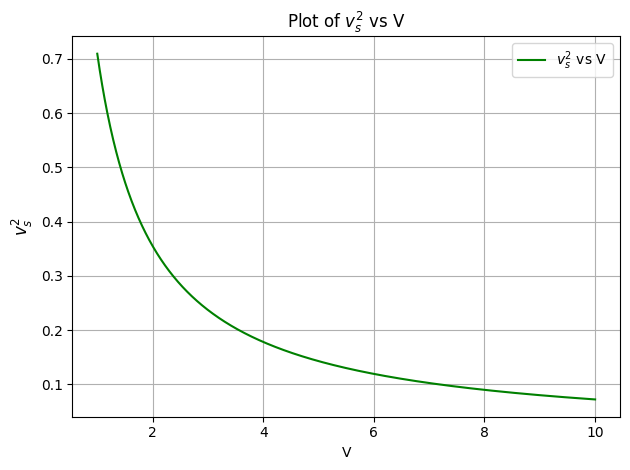}
        \caption{}
        \label{fig:v^2_MCJG}
    \end{subfigure}
    \begin{subfigure}[b]{0.45\textwidth}
        \includegraphics[width=1.07\textwidth]{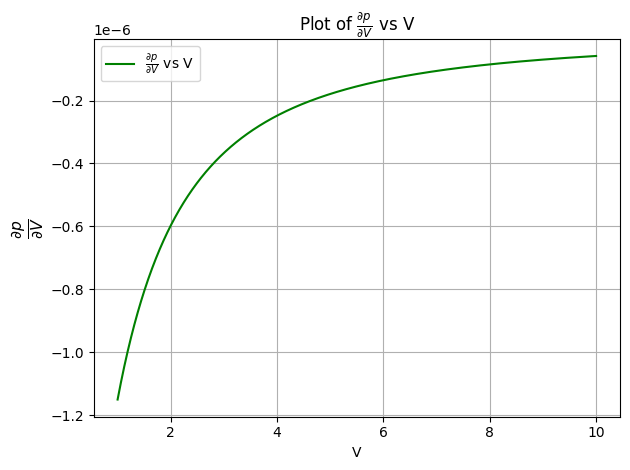}
        \caption{}
        \label{fig:dpdV_MCJG}
    \end{subfigure}
    \caption{Plot of square of the speed of sound $v_s^2$ and $\dfrac{\partial p}{\partial V}$ against volume $V$ for MCJG model.}
    \label{fig:v^2_dpdV_MCJG}
\end{figure}

\begin{figure}[htbp]
    \centering
    \begin{subfigure}[b]{0.45\textwidth}
        \includegraphics[width=1.03\textwidth]{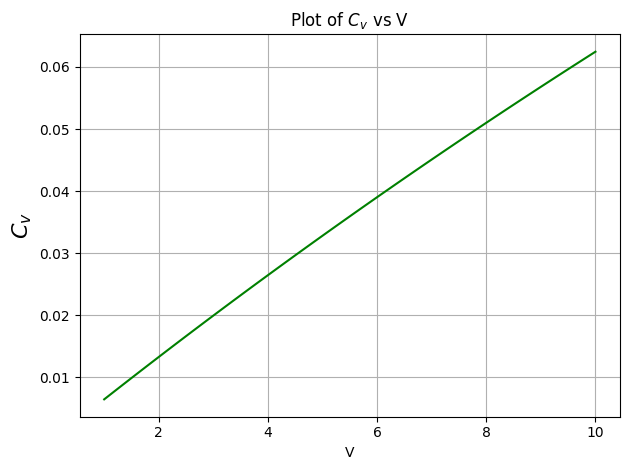}
        \caption{}
        \label{fig:Cv_MCJG}
    \end{subfigure}
    \begin{subfigure}[b]{0.45\textwidth}
        \includegraphics[width=1.03\textwidth]{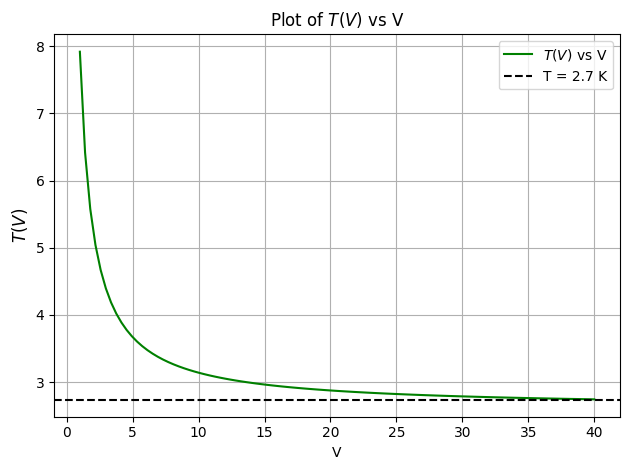}
        \caption{}
        \label{fig:T_MCJG}
    \end{subfigure}
    \caption{Plot of specific heat $C_V$ and temperature $T(V)$ against volume $V$ for MCJG model.}
    \label{fig:Cv_T_MCJG}
\end{figure}

In order to analyze the thermodynamic behavior, the following conditions must be considered:
\begin{itemize}
    \item During an adiabatic expansion, the pressure should decrease, implying $\left(\dfrac{\partial p}{\partial V}\right)_T < 0$.
    \item The heat capacity at constant volume must be positive, i.e., $C_V > 0$.
\end{itemize}
First we have calculated the value of $\left(\dfrac{\partial p}{\partial V}\right)_T$ for equation \eqref{p_MCJG},

\begin{equation}\label{delpdelV_MCJG}
    \left(\frac{\partial p}{\partial V}\right)_T = -\frac{BK\alpha\left(\frac{V_0}{V}\right)^{(A+1)(\alpha+1)}(A+1)}{V\left(K\left(\frac{V_0}{V}\right)^{(A+1)(\alpha+1)}+1\right)X}\quad,
\end{equation}
\noindent where $X$ is defined as
\begin{equation}\label{X_delpdelV_MCJG}
    X = \left[K\left(\frac{V_0}{V}\right)^{(A+1)(\alpha+1)}+\left(\frac{V_0}{V}\right)^{(A+1)(\alpha+1)}+1\right]
\left[\frac{B\left(K\left(\frac{V_0}{V}\right)^{(A+1)(\alpha+1)}+\left(\frac{V_0}{V}\right)^{(A+1)(\alpha+1)}+1\right)}
{(A+1)\left(K\left(\frac{V_0}{V}\right)^{(A+1)(\alpha+1)}+1\right)}\right]^{\frac{\alpha}{\alpha+1}}.
\end{equation}
From the above expression it is readily inferred that $B \neq 0$, because if so then $\left(\dfrac{\partial p}{\partial V}\right)_T = 0$ everywhere irrespective of the volume and other parameters. The figure.~\ref{fig:dpdV_MCJG} shows that the pressure reduces as the universe expands adiabatically. This imposes that the pressure of MCJG model defined in \eqref{MCJG_EoS} can attain any value without any constraint.
\\Now, To obtain $C_V$ first we have to calculate $T$ using the relation,
\begin{equation}\label{T_formula_MCJG}
    T = \dfrac{\partial U}{\partial S} = \Big(\dfrac{\partial U}{\partial C}\Big) \Big(\dfrac{\partial C}{\partial S}\Big).
\end{equation}
The consideration of parameters \(A\) and \(C\) as entropy-independent constants presents a theoretical challenge in the thermodynamic framework. Under such conditions, the system would exhibit a temperature reduction to zero, regardless of the gas's volumetric or pressure conditions. This would result in the \(T = 0\) isotherm simultaneously functioning as an isentropic pathway where \(S = \text{const}\). Such behavior generates a fundamental inconsistency with thermodynamic principles, specifically contradicting the third law. Therefore, a comprehensive examination of the MCJG thermodynamic stability necessitates the assumption that at minimum, one parameter must exhibit entropy \(S\) dependence. 
Now, the Internal Energy of the system is given as,
\begin{equation}\label{U_original_MCJG}
    U = V\left( \dfrac{B \left(1 - \dfrac{1}{1 - K - (\frac{B}{C})^{(A+1)} V^{(A+1)(\alpha+1)}}\right)}{1 + A} \right)^{\frac{1}{1 + \alpha}}.  
\end{equation}
Now, analysing the dimensions, the equation~\eqref{U_original_MCJG} gives,
\begin{equation}\label{C_definition_MCJG}
    C = (\tau S)^{(1+\alpha)},
\end{equation}
where, we have used $U=TS$. Here, $\tau$ is the constant with the dimension of temperature. Taking the derivative of the above equation with respect to S, we obtain
\begin{equation}\label{delC_delS_MCJG}
    \left( \frac{\partial C}{\partial S} \right)=(S \tau)^{\alpha + 1} \frac{\alpha + 1}{S}\quad.
\end{equation}
So, the expression of $T$ becomes,

\begin{equation}\label{T(S)_MCJG}
    T = \frac{V^{(A + 1)(\alpha + 1) + 1} \left( B (S \tau)^{-\alpha - 1} \right)^{A + 1} \left[ \frac{B \left( K + V^{(A + 1)(\alpha + 1)} \left( B (S \tau)^{-\alpha - 1} \right)^{A + 1} \right)}{(A + 1) \left( K + V^{(A + 1)(\alpha + 1)} \left( B (S \tau)^{-\alpha - 1} \right)^{A + 1} - 1 \right)} \right]^{\frac{1}{\alpha + 1}} (A + 1)}
{S \left[ K + V^{(A + 1)(\alpha + 1)} \left( B (S \tau)^{-\alpha - 1} \right)^{A + 1} \right] \left[ K + V^{(A + 1)(\alpha + 1)} \left( B (S \tau)^{-\alpha - 1} \right)^{A + 1} - 1 \right]} \quad.
\end{equation}
To determine whether the specific heat at constant volume is positive, we treat specific heat as a function of temperature and entropy, expressed as follows:

\begin{equation}\label{Cv_definition}
    C_v = T\left(\dfrac{\partial S}{\partial T}\right)_V.
\end{equation}
Now, we obtain the $\left(\frac{\partial S}{\partial T}\right)$ from the expression of $T$ and multiplying the same with $T$ we get,
\begin{equation}\label{Cv_MCJG}
    C_V = \frac{SV^{\beta+1}\Gamma^{A+1}(A+1)(K+\zeta)(K+\zeta-1)}
{V^{\beta+1}\Gamma^{A+1}(K+\zeta)[-\xi(K+\zeta-1) + V^{\beta}\Gamma^{A+1}(A+1)^2(2\alpha+3+\frac{1}{K+\zeta})]} \quad,
\end{equation}

\noindent where:

\begin{equation*}
    \beta = (A+1)(\alpha+1), \quad \Gamma = (B(S\tau)^{-\alpha-1})^{A+1}, \quad \zeta = V^{\beta}\Gamma, \quad \xi = A + \beta + 1 = (A+1)(2\alpha+3).
\end{equation*}
The positive value of specific heat at constant volume $C_V$ is shown in the fig.~\ref{fig:Cv_MCJG}. This, together with the previous conditions of $\left(\dfrac{\partial p}{\partial V}\right)_T<0$ ensure that MCJG is thermodynamically stable. Again, the temperature is given as,
\begin{equation}\label{T_definition}
    T = \dfrac{p+\rho}{S}\quad,
\end{equation}
putting the values of $p$ and $\rho$
\begin{equation}\label{T(V)_MCJG}
    T = -\frac{
    \dfrac{B K}{\left[B \left( \frac{(V_0/V)^{(A + 1)(\alpha + 1)}}{K (V_0/V)^{(A + 1)(\alpha + 1)} + 1} + 1 \right) / (A + 1)\right]^{\frac{\alpha}{\alpha + 1}} }
    - \left( \frac{B (2 K (V_0/V)^{(A + 1)(\alpha + 1)} + 1)}{A + 1} \right)^{\frac{1}{\alpha + 1}}
}{S}\quad.
\end{equation}
From this we can obtain another expression of $T$ as a function of volume $V$. We have plotted this expression of $T$ against $V$ in the following curve. It shows initially for small volume temperature is very large, as the universe expands the MCJG cools down and reaches to its current observed value which is $T \approx 2.7K$. This is visible in fig.~\ref{fig:T_MCJG}.

\subsection{Reduced Parameters}
From the integrability condition of a thermodynamic system it can be obtained that,
\begin{equation}\label{dp_dT_definition}
    \frac{dp}{dT} = \frac{p + \rho }{T}\quad.
\end{equation}
By solving the above equation, we get the pressure as a function of temperature as,
\begin{equation}\label{p(T)_MCJG}
    p(T) = -\left[(BK)^{\frac{1}{\alpha}}-\left(\frac{T}{T^*}\right)^{\frac{\alpha+1}{\alpha}}\right]^{\frac{\alpha}{\alpha + 1}}.
\end{equation}
Here, \( T^* \) represents an integration constant, with the condition \( 0 < T < T^* \), indicating that \( T^* \) is the highest temperature that the gas can achieve. Now let us assume the initial values of this system are given by $V = V_i$, $p = p_i$, $\rho = \rho_i$, $T = T_i$.
Now, from the equation~\eqref{rho_MCJG}, we can find the expression of the arbitrary parameter $C$ by using the initial values and is given as,
\begin{equation}\label{C/B_relation_MCJG}
    \left(\frac{C}{B}\right)^{(1+A)} = \left[\left(\frac{A+1}{B}\right)\rho_i^{1+\alpha} - 1 \right]\left(\frac{1}{2K}\right)V_i^{(1+A)(1+\alpha)}.
\end{equation}
Now, with this value, we can now derive the pressure and energy density in terms of the initial values.
\begin{equation}\label{rho_rhoi_MCJG}
    \rho = \rho_i \left[\frac{B}{(A+1)\rho_i^{1+\alpha}}+ \left(1-\frac{B}{(A+1)\rho_i^{1+\alpha}}\right)\left(\frac{V_i}{V}\right)^{(1+A)(1+\alpha)}\right]^{\frac{1}{1+\alpha}}\quad,
\end{equation}
\begin{equation}\label{p_pi_MCJG}
    p = -BK \left[\rho_i \left(\frac{B}{(A+1)\rho_i^{1+\alpha}}+ \left(1-\frac{B}{(A+1)\rho_i^{1+\alpha}}\right)\left(\frac{V_i}{V}\right)^{(1+A)(1+\alpha)}\right)^{\frac{1}{1+\alpha}}\right]^{-\alpha}\quad.
\end{equation}
We will now define the values of reduced parameters as, 
\begin{equation}\label{reduced_params_MCJG}
    \epsilon = \dfrac{\rho}{\rho_i}, \quad v = \dfrac{V}{V_i}, \quad P = \dfrac{p}{\left(\frac{B}{A+1}\right)^{-\frac{\alpha}{1+\alpha}}}, \quad \gamma = \dfrac{B}{(A+1)\rho_i^{1+\alpha}}, \quad t = \dfrac{T}{T_i}, \quad t^* = \dfrac{T^*}{T_i}.
\end{equation}
Now, using these definitions, our system variables can be represented as,
\begin{equation}
\epsilon = \left[\gamma + \dfrac{1-\gamma}{v^{(1+A)(1+\alpha)}}\right]^{\frac{1}{1+\alpha}},
\label{rho_reduced_MCJG}
\end{equation}

\begin{equation}
P = -BK \gamma^{\frac{\alpha}{1+\alpha}} \cdot \left[\gamma + \dfrac{1-\gamma}{v^{(1+A)(1+\alpha)}}\right]^{-\frac{\alpha}{1+\alpha}},
\label{p_reduced_MCJG}
\end{equation}

\begin{equation}
p(T) = -\left[(BK)^{\frac{1}{\alpha}}-\left(\frac{t}{t^*}\right)^{\frac{\alpha+1}{\alpha}}\right]^{\frac{\alpha}{\alpha + 1}}.
\label{p(T)_reduced_MCJG}
\end{equation}
By definition, when \( p = p_i \), \( V = V_i \), and \( T = T_i \), we have \( t = 1 \) and \( v = 1 \). Therefore, by equating \( p(T) \) and \( P \), we obtain:
\begin{equation}\label{p(T)equalP_MCJG}
    \left[(BK)^{\frac{1}{\alpha}}-\left(\frac{1}{t^*}\right)^{\frac{\alpha+1}{\alpha}}\right]^{\frac{\alpha}{\alpha + 1}}  = BK \gamma^{\frac{\alpha}{\alpha+1}}.
\end{equation}
The solution to this equation yields the value of \( t^* \) as follows:
\begin{equation}\label{t*_value_MCJG}
    t^* = \left(\dfrac{1}{BK}\right)^{\frac{1}{1+\alpha}}\left[\dfrac{1}{1-\gamma}\right]^{\frac{\alpha}{\alpha+1}},
\end{equation}
and, the value of $\gamma$ can be given as,
\begin{equation}\label{gamma_expression_MCJG}
    \gamma = 1 - (BK)^{-\frac{1}{\alpha}}\left(t^*\right)^{-\frac{\alpha+1}{\alpha}}.
\end{equation}
In our analysis, we consider two critical temperature points in the cosmological timeline: the theoretical peak temperature of the Modified Chaplygin-Jacobi Gas, denoted as \( T^* \), which is estimated at \( 10^{32} \) units, coinciding with temperature conditions during the Planck epoch. This is contrasted with the contemporary cosmic microwave background temperature, measured at \( T_i = 2.7 \) units. From these boundary conditions, we can proceed with our calculations.
\begin{equation}\label{gamma_value_MCJG}
    \gamma = 1 - (BK)^{-\frac{1}{\alpha}}\left(10^{32}\right)^{-\frac{\alpha+1}{\alpha}} \approx 1\quad.
\end{equation}
Consequently, in this scenario, the dark energy density of the universe permeated with Modified Chaplygin-Jacobi gas is nearly equal to its limiting value, given by \(\left(\dfrac{B}{A+1}\right)^{\frac{1}{\alpha+1}}\).
The internal energy of the system becomes,
\begin{equation}\label{U_modified_MCJG}
    U = \dfrac{(BK)^{\frac{1}{\alpha}}V}{\left[(BK)^{\frac{1}{\alpha}}-\left(\frac{T}{T^*}\right)^{\frac{\alpha+1}{\alpha}}\right]^{\frac{1}{\alpha + 1}}}\quad.
\end{equation}
Now, using the thermodynamic relation,
\begin{equation}\label{delU_delV_relation}
    \dfrac{\partial U}{\partial V}= T \dfrac{\partial p}{\partial T} - p \quad,
\end{equation}
we obtain, 
\begin{equation}\label{T*_finding equation_MCJG}
    \dfrac{(BK)^{\frac{1}{\alpha}}}{\left[(BK)^{\frac{1}{\alpha}}-\left(\frac{T}{T^*}\right)^{\frac{\alpha+1}{\alpha}}\right]^{\frac{1}{\alpha + 1}}} = \dfrac{BK}{\rho^\alpha}\left[\frac{T}{p} \frac{dp}{dT} - 1 \right]\quad.
\end{equation}
Solving this equation we can obtain the value of $T^*$.

\section{Thermodynamic Analysis of Modified Chaplygin-Abel Gas}\label{Sect:MCAG}

\begin{figure}[htbp]
    \centering
    \begin{subfigure}[b]{0.45\textwidth}
        \includegraphics[width=1.03\textwidth]{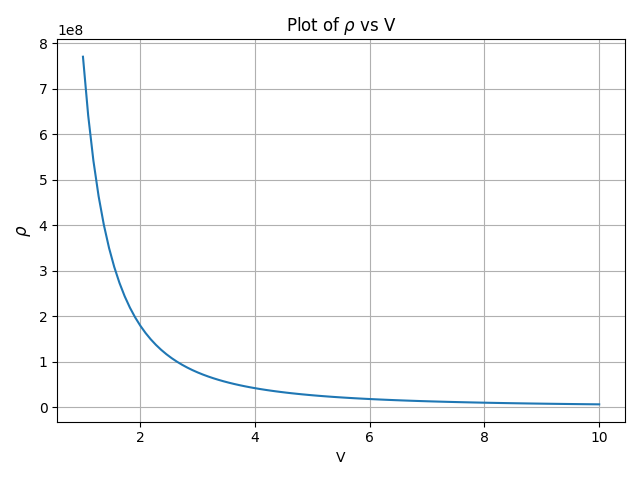}
        \caption{}
        \label{fig:rho_MCAG}
    \end{subfigure}
    \begin{subfigure}[b]{0.45\textwidth}
        \includegraphics[width=1.1\textwidth]{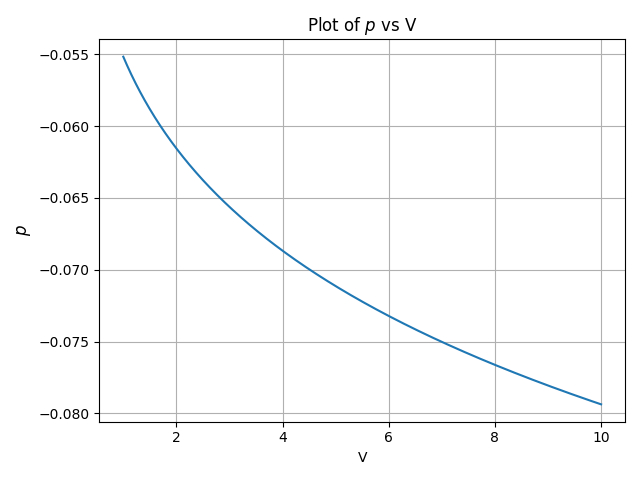}
        \caption{}
        \label{fig:p_MCAG}
    \end{subfigure}
    \caption{Plot of energy density $\rho$ and pressure $p$ against volume $V$ for MCAG model.}
    \label{fig:rho_p_MCAG}
\end{figure}

By substituting the hyperbolic function with the Abel elliptic function, we can derive the pressure-density relationship of the Modified Chaplygin-Abel gas as\cite{debnath_MCAG_MCJG},

\begin{equation}\label{MCAG_EoS}
    p = \left[(e^2 + 2c^2)(1 + A) - 1\right]\rho - \frac{c^2B}{\rho^\alpha} - \frac{(c^2 + e^2)(1 + A)^2\rho^{2 + \alpha}}{B} \quad.
\end{equation}
Here, $A$, $B$ are positive constant parameters, $0 \leq \alpha \leq 1$ and $c$, $e$ are real numbers. This is the pressure-density relation of ``Modified Chaplygin-Abel Gas''. The physical significances of $A$, $B$ and $\alpha$ parameters are detailed in Sect.~\ref{Sect:MCJG}. Though the elliptic parameters $e$, $c$ have no distinguished physical significance but they greatly affect the pressure-density relation of the modeled dark energy and henceforth have a deep impact on the evolution dynamics of the universe. The specific choice of these parameters can reduce the eq.~\eqref{MCAG_EoS} to another Chaplygin gas equation of state which we will see in the next paragraph. The observational constraints of their value is explicitly studied in \cite{debnath_MCAG_MCJG}.
Now, using equation~\eqref{rho_basic_definition} and \eqref{p_basic_definition} we get the Internal Energy as follows:
\begin{equation}\label{U_MCAG}
U = V \left[ \dfrac{Bc^2}{(1 + A)\left(c^2 + e^2\left(1 + \dfrac{1}{-1 + c^2 \exp((1 + A)(1 + \alpha)B e^2 d) V^{(1 + A)(1 + \alpha) e^2}}\right)\right)} \right]^{\frac{1}{1 + \alpha}}.
\end{equation}
Here, $d$ is an integration constant and it is function of entropy $(S)$ only[$d=d(S)$]. Now if we put $c=1$, $e=1$, $B=0$ and take the parameter $B$ to be large enough then the equation~\eqref{MCAG_EoS} reduces to,
\begin{equation}
    p = 2\rho - \frac{B}{\rho^\alpha}\quad.
\end{equation}
This is nothing but the pressure-density relation of Modified Chaplygin Gas with $A=2$~\cite{benaoum2022accelerated}. Now comparing the energy densities given by the MCG with that of MCAG we obtain,
\begin{equation}\label{MCAG_MCG_equivalence_condition}
    e^{(1+\alpha)Bd} = -\dfrac{B}{2D}\quad.
\end{equation}
Here, \( D \) is an arbitrary parameter that depends solely on the entropy \( S \).
Now, using the above relation the energy density takes the form as, 
\begin{equation}
    \rho = \left[ \dfrac{B c^2}{(A + 1) \left( c^2 + e^2 \left( 1 + \frac{1}{V^{e^2 (A + 1) (\alpha + 1)} c^2 \left( -\frac{B}{2D} \right)^{e^2 (A + 1)} - 1} \right) \right)} \right]^{\frac{1}{\alpha + 1}}.
\end{equation}
Again if we take further approximations the energy density equation reduces to the following,
\begin{equation}\label{rho_MCAG}
    \rho = \left[ \frac{B}{(A + 1)} \left( 1 +\frac{e^2}{c^2} \left(\frac{V_0}{V}\right)^{e^2 (A + 1) (\alpha + 1)} \right) \right]^{\dfrac{1}{\alpha + 1}} .
\end{equation}
Here, we have used $$V_0 ^{(1+\alpha)} = \dfrac{2D}{B}\quad.$$  From the equation \eqref{rho_MCAG} it is clearly obtainable that at very low volume i.e. for $V \to 0$, the energy density becomes very high. Alternatively, at very large volume i.e for $V \to \infty$  the energy represents the mixture of two fluids with one has constant energy density and other has a volume dependent energy density given as,

\begin{equation}\label{rho_high_volume_MCAG}
    \rho \approx \left (\frac{B}{A+1}\right)^{\frac{1}{1+\alpha}} + \frac{1}{1+\alpha}\left(\frac{e^2}{c^2}\right)\left(\frac{V_0}{V}\right)^{e^2 (A + 1) (\alpha + 1)}.
\end{equation}
Now, The corresponding pressure can be obtained as,
\begin{equation}\label{p_MCAG}
    p = -\dfrac{B c^2}{\left[ \frac{B }{A + 1} \left( 1 + \frac{e^2}{c^2} \left( \frac{V_0}{V} \right)^{e^2 (A + 1) (\alpha + 1)} \right) \right]^{\frac{\alpha}{\alpha + 1}}}\quad.
\end{equation}
The equation imparts that for small volume (and high energy density) pressure becomes,
\begin{equation}\label{p_small_volume_MCAG}
    p \approx -Bc^2 \left[\left(\frac{B}{A+1}\right)^{-\frac{\alpha}{1+\alpha}} \left(\frac{V}{V_0}\right)^{e^2 \alpha(1+A)}\right] \approx 0,
\end{equation}
and for large volume (low energy density) the pressure becomes a negative constant which resembles the cosmological constant scenario,
\begin{equation}\label{p_large_volume_MCAG}
    p \approx -Bc^2 \left[\frac{B}{A+1}\right]^{-\frac{\alpha}{\alpha+1}}.
\end{equation}
The similar behaviour of energy density and pressure is depicted in fig.\ref{fig:rho_MCAG} and \ref{fig:p_MCAG}.

\subsection{EoS Parameter}
The equation of state parameter is expressed as \( \omega = \dfrac{p}{\rho} \). By substituting the values of pressure and energy density, the resulting equation of state parameter for the Modified Chaplygin-Abel Gas is given by:

\begin{equation}\label{EoS_MCAG}
    \omega = -\frac{B c^2}{\left[ \frac{B \left( c^2 + e^2 \left( \frac{V_0}{V} \right)^{e^2 (A + 1) (\alpha + 1)} \right)}{c^2 (A + 1)} \right]^{\frac{1}{\alpha + 1}} \left[ \left( \frac{B \left( c^2 + e^2 \left( \frac{V_0}{V} \right)^{e^2 (A + 1) (\alpha + 1)} \right)}{c^2 (A + 1)} \right)^{\frac{1}{\alpha + 1}} \right]^\alpha}\quad.
\end{equation}
It represents a very small constant value of EoS parameter at small volume ($V \to 0$) and at large volume ($V \to \infty$) the EoS parameter becomes negative and volume dependent and is plotted in fig.~\ref{fig:omega_MCAG}. It is worthy to note that for the chosen values of $B=c=1$ and $V_0=A=0$, the EoS $\omega$ becomes $-1$ which is the dark energy equation of state parameter in the standard $\Lambda$CDM paradigm irrespective of the values of $\alpha$ and $e$.

\subsection{Decelerating Parameter}
\begin{figure}[htbp]
    \centering
    \begin{subfigure}[b]{0.45\textwidth}
        \includegraphics[width=\textwidth]{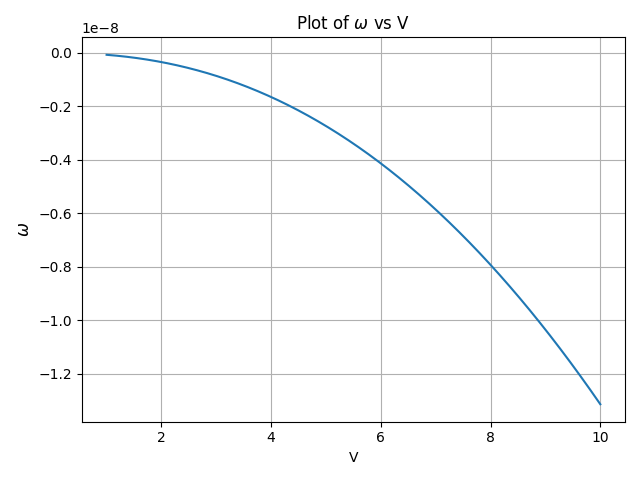}
        \caption{}
        \label{fig:omega_MCAG}
    \end{subfigure}
    \begin{subfigure}[b]{0.45\textwidth}
        \includegraphics[width=\textwidth]{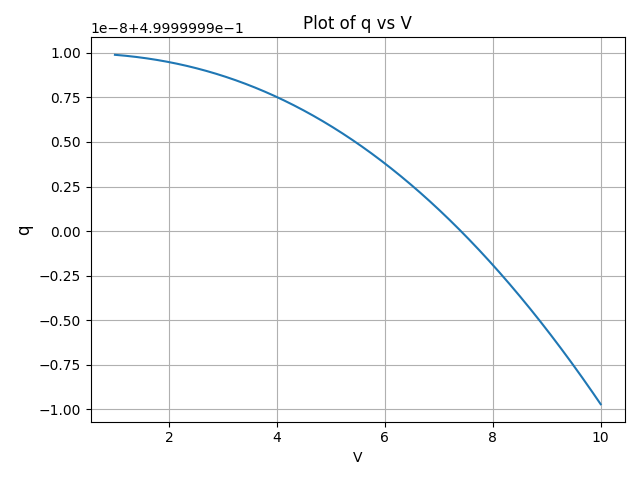}
        \caption{}
        \label{fig:q_MCAG}
    \end{subfigure}
    \caption{Graphical representation illustrating the relationship between the volume $V$ and two key cosmological indicators in the Modified Chaplygin-Abel gas framework: the equation of state coefficient $\omega$ and the deceleration factor $q$.}
    \label{fig:omega_q_MCAG}
\end{figure}

The deceleration parameter \( q \) indicates the rate of expansion of the universe. A negative value of \( q \) signifies that the universe is undergoing an accelerated expansion phase, while a positive value indicates a decelerating phase of expansion.
Using the definition from equation \eqref{q_definition}, we obtained
\begin{equation}\label{q_MCAG}
    q = -\frac{3 B c^2}{2 \left[ \frac{B \left( 1 + \frac{e^2}{c^2} \left( \frac{V_0}{V} \right)^{e^2 (A + 1) (\alpha + 1)} \right)}{(A + 1)} \right]^{\frac{1}{\alpha + 1}} \left[ \left( \frac{B \left( 1 + \frac{e^2}{c^2}\left( \frac{V_0}{V} \right)^{e^2 (A + 1) (\alpha + 1)} \right)}{(A + 1)} \right)^{\frac{1}{\alpha + 1}} \right]^\alpha} + 0.5\quad.
\end{equation}
Here, when the volume is small then $\frac{V}{V_0}$ term becomes small and less than $0.5$ making the deceleration parameter positive and for large volume  $\frac{V}{V_0}$ term dominates and becomes larger than $0.5$ making the deceleration parameter negative.
Therefore, negative values of the deceleration parameter suggest that the universe is undergoing an accelerated expansion phase at large scales. This transition can be observed in fig.~\ref{fig:q_MCAG}.

\subsection{Square Speed of Sound}
The speed of sound $v_s^2$ is a measure of stability of a model. If $v_s^2 > 0$ then the model is stable, otherwise it is unstable.
\newline By using the definition from subsect.~\ref{subsect:soundsquare} we have obtained,
\begin{equation}\label{v_square_MCAG}
    v_s^2 = \frac{B \alpha c^2}{\left[ \frac{B \left( 1 + \frac{e^2}{c^2} \left( \frac{V_0}{V} \right)^{e^2(A + 1) (\alpha + 1)} \right)}{ (A + 1)} \right]^{\frac{1}{\alpha + 1}} \left[ \left( \frac{B \left( 1 + \frac{e^2}{c^2} \left( \frac{V_0}{V} \right)^{e^2 (A + 1) (\alpha + 1)} \right)}{ (A + 1)} \right)^{\frac{1}{\alpha + 1}} \right]^\alpha} + (A + 1) \left( 2c^2 + e^2 \right) - 1 \quad.
\end{equation}
The figure.~\ref{fig:v^2_MCAG} shows that $v_s^2$ is positive throughout the expansion of the universe and hence classically stable under perturbation.

\subsection{Thermodynamic Stability}
\begin{figure}[htbp]
    \centering
    \begin{subfigure}[b]{0.45\textwidth}
        \includegraphics[width=1.03\textwidth]{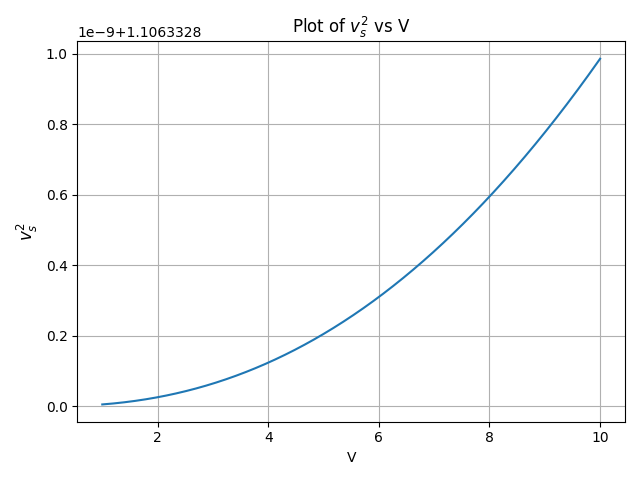}
        \caption{}
        \label{fig:v^2_MCAG}
    \end{subfigure}
    \begin{subfigure}[b]{0.45\textwidth}
        \includegraphics[width=1.07\textwidth]{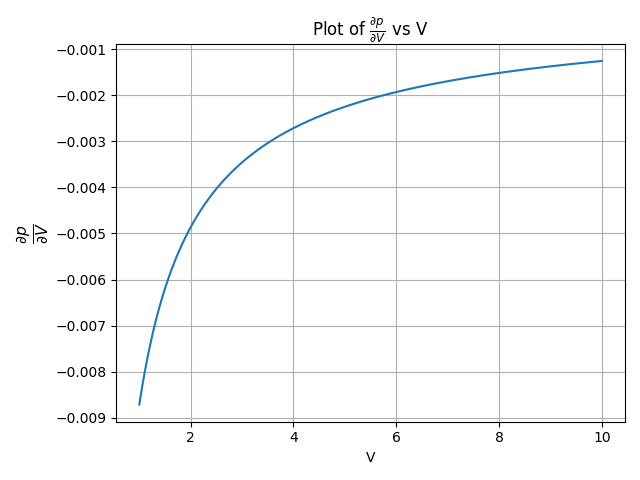}
        \caption{}
        \label{fig:dpdV_MCAG}
    \end{subfigure}
    \caption{Plot of square of the speed of sound $v_s^2$ and $\dfrac{\partial p}{\partial V}$ against volume $V$ for MCAG model.}
    \label{fig:v^2_dpdV_MCAG}
\end{figure}

\begin{figure}[htbp]
    \centering
    \begin{subfigure}[b]{0.45\textwidth}
        \includegraphics[width=1.03\textwidth]{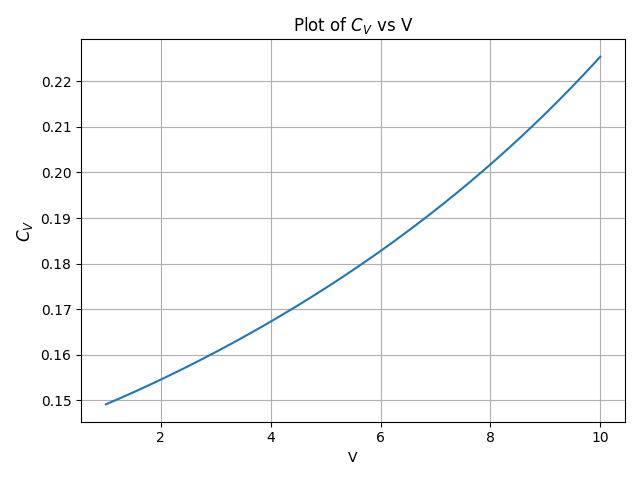}
        \caption{}
        \label{fig:Cv_MCAG}
    \end{subfigure}
    \begin{subfigure}[b]{0.45\textwidth}
        \includegraphics[width=1.03\textwidth]{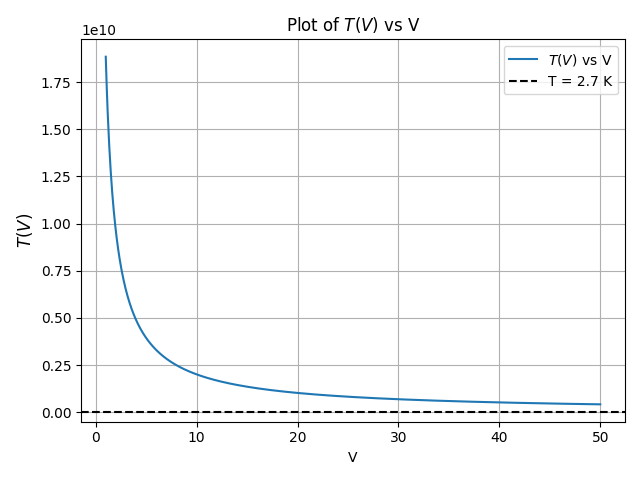}
        \caption{}
        \label{fig:T_MCAG}
    \end{subfigure}
    \caption{Plot of specific heat $C_V$ and temperature $T(V)$ against volume $V$ for MCAG model.}
    \label{fig:Cv_T_MCAG}
\end{figure}

To establish the conditions for thermodynamic stability, we need to take into account the following criteria:
\begin{itemize}
    \item During an adiabatic expansion, the pressure must decrease, indicating that $\left(\dfrac{\partial p}{\partial V}\right)_T < 0$.
    \item The heat capacity at constant volume should be positive, which means $C_V > 0$.
\end{itemize}
First we have calculated the value of $\left(\dfrac{\partial p}{\partial V}\right)_T$ for equation \eqref{p_MCAG},

\begin{equation}\label{delP_delV_MCAG}
    \left(\frac{\partial p}{\partial V}\right)_T = -\frac{B \alpha e^4 \left( \frac{V_0}{V} \right)^{e^2 (A + 1) (\alpha + 1)} (A + 1)}{V \left( 1 + \frac{e^2}{c^2} \left( \frac{V_0}{V} \right)^{e^2 (A + 1) (\alpha + 1)} \right) \left( \frac{B \left( 1 + \frac{e^2}{c^2} \left( \frac{V_0}{V} \right)^{e^2 (A + 1) (\alpha + 1)} \right)}{ (A + 1)} \right)^{\frac{\alpha}{\alpha + 1}}}\quad.
\end{equation}
From the above expression it is readily inferred that $B \neq 0$, because if so then $\left(\dfrac{\partial p}{\partial V}\right)_T= 0$ everywhere irrespective of the volume and other parameters. The figure.~\ref{fig:dpdV_MCAG} shows that the pressure reduces as the universe expands adiabatically. This imposes that the pressure of MCAG model defined in \eqref{MCAG_EoS} can attain any value without any constrain.
\\Now, To obtain $C_V$ first we have to calculate $T$ using the relation,
\begin{equation}\label{T_definition_MCAG}
    T = \dfrac{\partial U}{\partial S} = \Big(\dfrac{\partial U}{\partial D}\Big) \Big(\dfrac{\partial D}{\partial S}\Big).
\end{equation}
When examining the thermodynamic properties of the system, a critical consideration emerges regarding the parameters \(A\) and \(D\). Should these parameters be considered entropy-independent constants, the system exhibits a problematic behavior wherein the temperature approaches zero irrespective of the gas's pressure or volumetric state. This scenario creates a thermodynamically inconsistent situation where the zero-temperature isotherm (\(T = 0\)) becomes coincident with an isentropic pathway (\(S = \text{const}\)). Such behavior fundamentally conflicts with established thermodynamic principles, particularly the third law. Consequently, a rigorous analysis of the MCAG thermodynamic stability requires the postulation that entropy dependence exists in at least one of these parameters \(S\).
Now, the internal energy $U$ is given as 
\begin{equation}\label{U_original_MCAG}
    U = V \left( \frac{B c^2}{(A + 1) \left( c^2 + e^2 \left( 1 + \frac{1}{V^{e^2 (A + 1) (\alpha + 1)} c^2 \left( -\frac{B}{2D} \right)^{e^2 (A + 1)} - 1} \right) \right)} \right)^{\frac{1}{\alpha + 1}}.
\end{equation}
Now, using concepts of dimensional analysis and using $U =[T][S]$ equation \eqref{U_original_MCAG} gives,
\begin{equation}\label{D_expression_MCAG}
    D = \nu^{2(1 + \alpha)} (\tau S)^{-1 - \alpha},
\end{equation}
thus,
\begin{equation}\label{delD_delS_MCAG}
    \frac{\partial D}{\partial S} = \frac{\nu^{2\alpha + 2} (S \tau)^{-\alpha - 1} (-\alpha - 1)}{S}\quad.
\end{equation}
Now, we have put the value of $D$ from equation \eqref{D_expression_MCAG} and $\frac{\partial D}{\partial S}$ from equation \eqref{delD_delS_MCAG} in the expression of $T$ and also calculated $\frac{\partial U}{\partial D}$ from \eqref{U_original_MCAG}. Doing this, we get $T$ as a function of entropy $S$. From this we have calculated $\frac{\partial T}{\partial S}$ by differentiating T with respect to $S$.
\newline Finally we have derived the heat capacity for constant volume $C_V$  using the relation $C_V = T\Big(\dfrac{\partial S}{\partial T}\Big)$ and obtained,
\begin{align}\label{Cv_MCAG}
  C_V &= \dfrac{R \left( Z - X E Y \right) \left( -Z + X E Y + X O Y \right)}
{\left( F \left( Z - X E Y \right) (O N - 1) \left( -Z + X E Y + X O Y \right) 
+ F^2 E O Y N \left( -Z + X E Y + X O Y \right) \right)} \\ \nonumber
&- F^2 O Y \left( Z - X E Y \right) S (E + O) \\ \nonumber
&- F^2 O Y N \left( E \left( -Z + X E Y + X O Y \right) + \left( Z - X E Y \right)(E + O) \right)\quad, 
\end{align}
where, we have assumed,
\begin{align*}
    X &= V^{e^2(A+1)(\alpha+1)}, \quad Y = (-B \nu^{-2\alpha-2} (S \tau)^{\alpha+1})^{e^2(A+1)}, \quad Z = 2^{e^2(A+1)}, \\
    E &= c^2, \quad F = V^{e^2(A+1)(\alpha+1)+1}, \quad N = (A+1)(\alpha+1), \quad O = e^2.
\end{align*}
The positive value of specific heat at constant volume $C_V$ is shown in the fig.~\ref{fig:Cv_MCAG}. This, together with the previous conditions of $\left(\dfrac{\partial p}{\partial V}\right)_T<0$ ensure that MCAG is thermodynamically stable.
Again, the temperature can also be defined as in eq.~\eqref{T_definition}
Putting the values of $p$ and $\rho$ from eq.~\eqref{rho_MCAG} and \eqref{p_MCAG} in \eqref{T_definition} we get,
\begin{equation}\label{T(V)_MCAG}
    T = \left(\dfrac{1}{S}\right)\left[ \frac{-B c^2}{\left( \frac{B \left(1 + \frac{e^2}{c^2} \left(\frac{V_0}{V}\right)^{e^2(A + 1)(\alpha + 1)}\right)}{A+1} \right)^{\frac{\alpha}{\alpha + 1}}} + \left( \frac{B \left(1 + \frac{e^2}{c^2} \left(\frac{V_0}{V}\right)^{e^2(A + 1)(\alpha + 1)}\right)}{A+1} \right)^{\frac{1}{\alpha + 1}} \right] \quad.
\end{equation}
From this we can obtain another expression of $T$ as a function of volume $V$. We have plotted this expression of $T$ against $V$ in the following curve. It shows initially for small volume temperature is very large, as the universe expands the MCAG cools down and reaches to its current observed value which is $T \approx 2.7K$. This is visible in fig.~\ref{fig:T_MCAG}.

\subsection{Reduced Parameters}
From the integrability condition of a thermodynamic system \eqref{dp_dT_definition}, we get the pressure as a function of temperature as,
\begin{equation}\label{p(T)_MCAG}
    p(T) = -\left[(Bc^2)^{\frac{1}{\alpha}}-\left(\frac{T}{T^*}\right)^{\frac{\alpha+1}{\alpha}}\right]^{\frac{\alpha}{\alpha + 1}}.
\end{equation}
Here, \( T^* \) represents an integration constant, and the condition \( 0 < T < T^* \) indicates that \( T^* \) is the highest temperature achievable by the gas. Now, let us consider the initial values of this system to be \( V = V_i \), \( p = p_i \), \( \rho = \rho_i \), and \( T = T_i \).
Now, from the equation \eqref{rho_MCAG}, we can find the expression of the arbitrary parameter $D$ in terms of the initial values as,
\begin{equation}\label{D/B_expression_MCAG}
    \left(\frac{2D}{B}\right)^{e^2(1+A)} = \left[\left(\frac{A+1}{B}\right)\rho_i^{1+\alpha} - 1 \right]\left(\frac{c}{e}\right)^2V_i^{e^2(1+A)(1+\alpha)}.
\end{equation}
Now, utilizing this value, we can express the pressure and energy density in terms of the initial values.
\begin{equation}\label{rho_rhoi_MCAG}
    \rho = \rho_i \left[\frac{B}{(A+1)\rho_i^{1+\alpha}}+ \left(1-\frac{B}{(A+1)\rho_i^{1+\alpha}}\right)\left(\frac{V_i}{V}\right)^{e^2(1+A)(1+\alpha)}\right]^{\frac{1}{1+\alpha}},
\end{equation}
\begin{equation}\label{p_pi_MCAG}
    p = -Bc^2 \left[\rho_i \left(\frac{B}{(A+1)\rho_i^{1+\alpha}}+ \left(1-\frac{B}{(A+1)\rho_i^{1+\alpha}}\right)\left(\frac{V_i}{V}\right)^{e^2(1+A)(1+\alpha)}\right)^{\frac{1}{1+\alpha}}\right]^{-\alpha}.
\end{equation}
We will now define the values of reduced parameters as,
\begin{equation}\label{reduced_params_MCAG}
    \epsilon = \dfrac{\rho}{\rho_i}, \quad v = \dfrac{V}{V_i}, \quad P = \dfrac{p}{\left(\frac{B}{A+1}\right)^{-\frac{\alpha}{1+\alpha}}}, \quad \gamma = \dfrac{B}{(A+1)\rho_i^{1+\alpha}}, \quad t = \dfrac{T}{T_i}, \quad t^* = \dfrac{T^*}{T_i}.
\end{equation}
Now, using these definitions, our system variables can be represented as,

\begin{align}
\epsilon &= \left[\gamma + \dfrac{1-\gamma}{v^{e^2(1+A)(1+\alpha)}}\right]^{\frac{1}{1+\alpha}}, \label{reduced_rho_MCAG} \\
P &= -Bc^2 \gamma^{\frac{\alpha}{1+\alpha}} \cdot \left[\gamma + \dfrac{1-\gamma}{v^{e^2(1+A)(1+\alpha)}}\right]^{-\frac{\alpha}{1+\alpha}}, \label{reduced_p_MCAG} \\
p(T) &= -\left[(Bc^2)^{\frac{1}{\alpha}}-\left(\frac{t}{t^*}\right)^{\frac{\alpha+1}{\alpha}}\right]^{\frac{\alpha}{\alpha + 1}}. \label{reduced_p(T)_MCAG}
\end{align}
By the definition, at $p = p_i$, $V = V_i$ and $T=T_i$, we have $t = 1$ and $v=1$. Thus, equating $p(T)$ and $P$ we get, 
\begin{equation}\label{p_p(T)_equation}
    \left[(Bc^2)^{\frac{1}{\alpha}}-\left(\frac{1}{t^*}\right)^{\frac{\alpha+1}{\alpha}}\right]^{\frac{\alpha}{\alpha + 1}}  = Bc^2 \gamma^{\frac{\alpha}{\alpha+1}}.
\end{equation}
The solution to this equation provides the value of \( t^* \) as follows:

\begin{equation}\label{t*_MCAG}
    t^* = \left(\dfrac{1}{Bc^2}\right)^{\frac{1}{1+\alpha}}\left[\dfrac{1}{1-\gamma}\right]^{\frac{\alpha}{\alpha+1}},
\end{equation}
and, the value of $\gamma$ can be given as,
\begin{equation}\label{gamma_expression_MCAG}
    \gamma = 1 - (Bc^2)^{-\frac{1}{\alpha}}\left(t^*\right)^{-\frac{\alpha+1}{\alpha}}.
\end{equation}
Our analysis establishes two crucial temperature benchmarks for the Modified Chaplygin-Abel Gas: an upper bound temperature \(T^*\), which reaches \(10^{32}\) units corresponding to conditions during the Planck epoch, and the current cosmic background temperature \(T_i\), measured at \(2.7\) units. These boundary conditions enable us to perform the following derivation:

\begin{equation}\label{gamma_value_MCAG}
    \gamma = 1 - (Bc^2)^{-\frac{1}{\alpha}}\left(10^{32}\right)^{-\frac{\alpha+1}{\alpha}} \approx 1\quad.
\end{equation}
In this scenario, the energy density of the universe composed of MCAG tends toward its limiting value expressed as \(\left(\dfrac{B}{A+1}\right)^{\frac{1}{\alpha+1}}\).
The internal energy of the system becomes,
\begin{equation}\label{U_modified_MCAG}
    U = \dfrac{(Bc^2)^{\frac{1}{\alpha}}V}{\left[(Bc^2)^{\frac{1}{\alpha}}-\left(\frac{T}{T^*}\right)^{\frac{\alpha+1}{\alpha}}\right]^{\frac{1}{\alpha + 1}}}\quad.
\end{equation}
Now, using the thermodynamic relation~\eqref{delU_delV_relation}, we obtain, 
\begin{equation}\label{T*_value_equation_MCAG}
    \dfrac{(Bc^2)^{\frac{1}{\alpha}}}{\left[(Bc^2)^{\frac{1}{\alpha}}-\left(\frac{T}{T^*}\right)^{\frac{\alpha+1}{\alpha}}\right]^{\frac{1}{\alpha + 1}}} = \dfrac{Bc^2}{\rho^\alpha}\left[\frac{T}{p} \frac{dp}{dT} - 1 \right]\quad.
\end{equation}
Solving this equation we can obtain the value of $T^*$.

\section{Observational Analysis}\label{Sect:Observational Study}
\begin{figure}[htbp]
    \centering
    \includegraphics[width=0.9\textwidth]{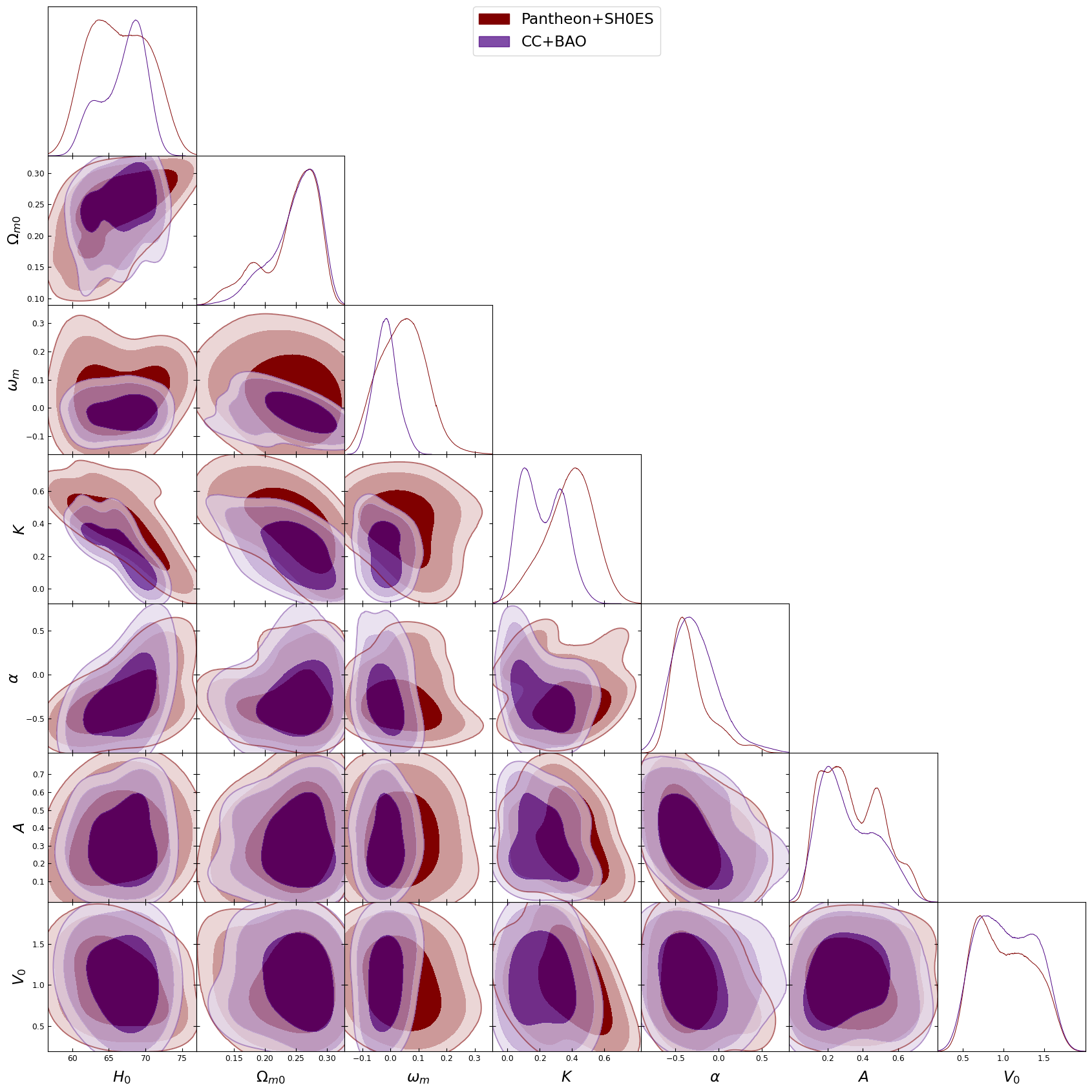}
    \caption{Posterior distribution at 1$\sigma$ and 2$\sigma$ confidence level for MCJG Model.}
    \label{fig:getdist_MCJG}
\end{figure}

\begin{figure}[htbp]
    \centering
    \includegraphics[width=0.9\textwidth]{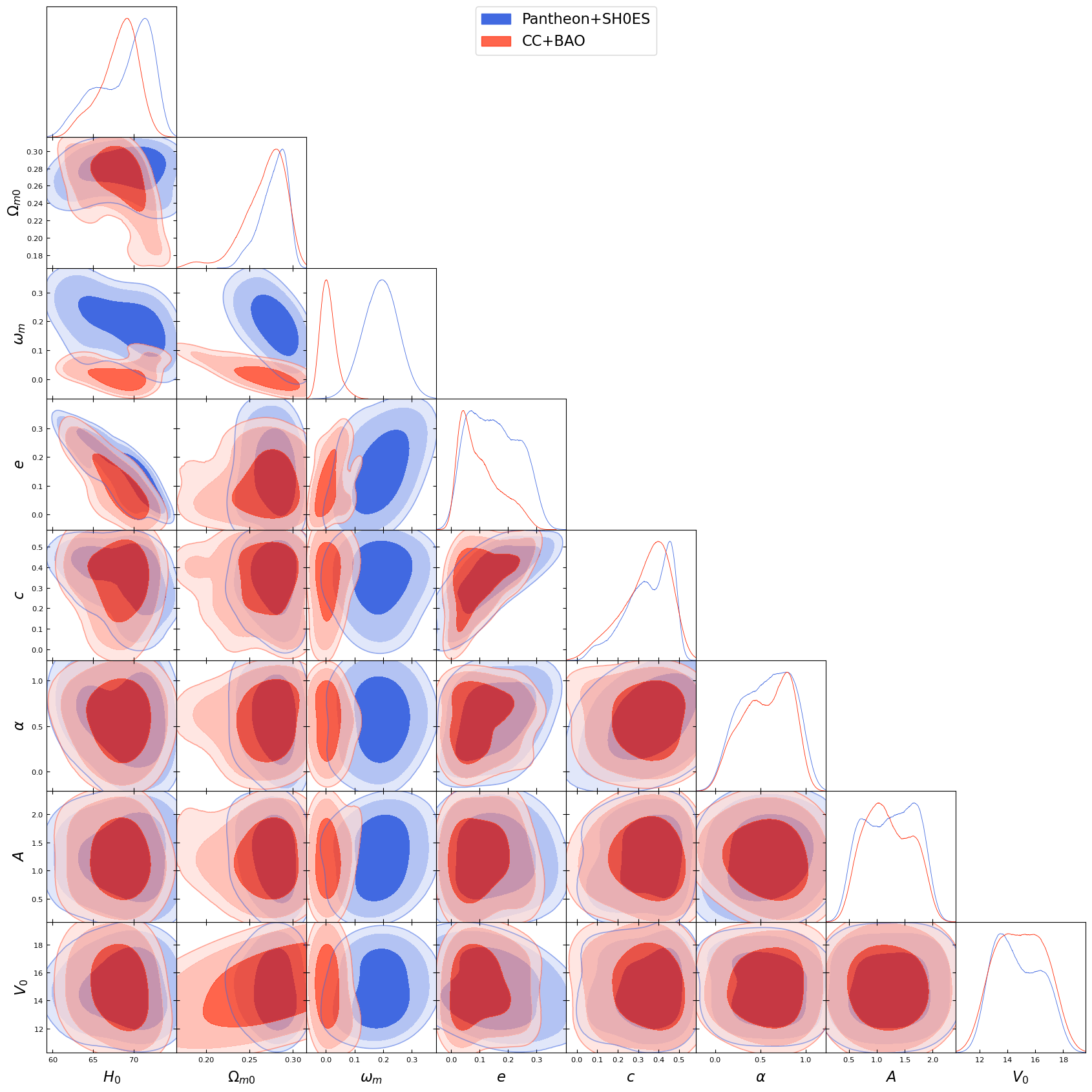}
    \caption{Posterior distribution at 1$\sigma$ and 2$\sigma$ confidence level for MCAG Model.}
    \label{fig:getdist_MCAG}
\end{figure}

In this section of the paper, we outline the observational datasets and the methodology employed to constrain the free parameters of the Modified Chaplygin-Jacobi Gas and Modified Chaplygin-Abel Gas models. We utilized the Markov Chain Monte Carlo (MCMC) algorithm, specifically employing the \texttt{emcee} package \cite{foreman2013emcee}. The likelihood function was constructed assuming Gaussian errors in the observational data, and uniform priors were applied to the model parameters based on physically motivated boundaries.
\\The MCMC simulations were executed with a total of 10,000 iterations, following an initial burn-in phase of 100 steps to ensure proper convergence of the chains. We employed 100 walkers to efficiently explore the parameter space. The convergence of the MCMC chains was assessed using the Gelman-Rubin statistic, with values of \( R < 1.1 \) for all parameters signifying adequate convergence. The optimal parameter values obtained from the analysis are presented in Table~\ref{tab:MCJG_MCAG_best_fit_params}, while the posterior distributions at the \( 1\sigma \) and \( 2\sigma \) confidence levels are illustrated in Figures~\ref{fig:getdist_MCJG} and \ref{fig:getdist_MCAG}.
\\The derivation of the Friedmann equations in cosmological theory begins with an examination of the Einstein field equations in the context of a Friedmann-Lemaître-Robertson-Walker (FLRW) spacetime geometry. The fundamental spacetime interval in this framework is characterized by:
\[
ds^2 = -dt^2 + a^2(t) \left[ \frac{dr^2}{1 - kr^2} + r^2 \left( d\theta^2 + \sin^2 \theta \, d\phi^2 \right) \right]\quad.
\]
Here, the temporal evolution of the universe is encoded in the scale factor $a(t)$, while $k$ represents the spatial geometry parameter. The cornerstone of general relativity, the Einstein field equations, takes the form:
\[
G_{\mu\nu} = 8\pi G T_{\mu\nu}\quad,
\]
where the geometric properties of spacetime are encoded in the Einstein tensor $G_{\mu\nu}$, and the matter-energy content is described by the stress-energy tensor $T_{\mu\nu}$.
When we incorporate the FLRW metric into these equations, we obtain the foundational Friedmann equations:
\begin{align}
3H^2 &=\kappa(\rho_m+\rho_{de}), \label{friedmann_1} \\
3H^2+2\dot{H} &=-\kappa(p_m+p_{de}). \label{friedmann_2}
\end{align}
In these expressions, $H = \frac{\dot{a}}{a}$ represents the cosmic expansion rate (Hubble parameter), $\kappa = \frac{8 \pi G}{c}$ is the gravitational coupling constant, while $\rho$ and $p$ denote the energy density and pressure respectively. For analytical simplicity, we adopt a flat universe model with $k = 0$ and normalize $\kappa = 1$.
\\The relationship between matter and dark energy density parameters emerges from the density parameter formulation:
\[
\Omega_{\text{total}} = \Omega_m + \Omega_{de} = \frac{\rho}{\rho_{\text{crit}}}\quad,
\]
where the critical density $\rho_{\text{crit}} = \frac{3H^2}{8\pi G}$ serves as a fundamental cosmic parameter. In the context of a flat universe, the constraint $\Omega_{\text{total}} = 1$ naturally emerges.

\begin{equation}\label{Omegade_constraint}
    \Omega_m + \Omega_{de} = 1.
\end{equation}
The matter and dark energy components each satisfy their respective continuity equations, as there is no interaction between the matter sector and the geometry. These equations are expressed as:

\begin{equation}\label{mattercontinuity}
\dot{\rho}_{m} + 3H(\rho_{m} + p_{m}) = 0,
\end{equation}
and

\begin{equation}\label{DEcontinuity}
\dot{\rho}_{de} + 3H(\rho_{de} + p_{de}) = 0.
\end{equation}
The solution of the Eq.~\eqref{mattercontinuity} is written as:

\begin{equation}\label{rho_mvalue}
\rho_{m} = \rho_{m0} (1 + z)^{3(1 + \omega_{m})},
\end{equation}
where $\rho_{m0}$ represents the present-day matter density. Additionally, using the definition $\omega_m = \dfrac{p_m}{\rho_m}$, the pressure is given by:

\begin{equation}\label{p_mvalue}
p_{m} = \omega_{m} \rho_{m0} (1 + z)^{3(1 + \omega_{m})}.
\end{equation}
In this context, \( \omega_m \) represents the constant equation of state (EoS) parameter for the matter component.
Using Eq.~\eqref{friedmann_1}, the Hubble parameter can be expressed in terms of \( \Omega_m \) and \( \Omega_{de} \) as follows:

\begin{equation}\label{Hubble_General_Relation}
    H(z) = H_0 \left[\Omega_{m0}(1+z)^{3(1+\omega_m)}+\rho_{de}\right]^{\frac{1}{2}},
\end{equation}
where $H_0$ is the present value of the Hubble parameter and $z$ is the redshift.
\\Now, we will use the values of the dark energy density $\rho_{DE}$ as given in the equations~\eqref{rho_MCJG} and \eqref{rho_MCAG} in the eq.~\eqref{Hubble_General_Relation}. Thus we obtain the Hubble parameter for MCJG and MCAG model as follows:
\begin{gather}
    H(z)_{\text{MCJG}} = H_0 \left[\Omega_{m0}(1+z)^{3(1+\omega_m)}+(1-\Omega_{m0})\left(1 + 2K\left(\dfrac{V_0}{V}\right)^{(\alpha+1)(A+1)}\right)^{\frac{1}{1 + \alpha}}\right]^{\dfrac{1}{2}}, \label{H_MCJG} \\
    H(z)_{\text{MCAG}} = H_0 \left[\Omega_{m0}(1+z)^{3(1+\omega_m)}+(1-\Omega_{m0})\left( 1 +\frac{e^2}{c^2} \left(\frac{V_0}{V}\right)^{e^2 (A + 1) (\alpha + 1)} \right)^{\frac{1}{\alpha + 1}} \right]^{\dfrac{1}{2}}. \label{H_MCAG}
\end{gather}
Here, \( \Omega_{m0} = \frac{\rho_{m0}}{3H_0^2} \) represents the present value of the matter density parameter, where \( \rho_{m0} \) denotes the matter density in the current epoch, and \( \omega_m \) is the equation of state (EoS) parameter for matter. The dark energy density parameter is expressed as \( \Omega_{d0} = \frac{1}{3H_0^2}\left(\frac{B}{A+1}\right)^{\frac{1}{\alpha + 1}} \). Additionally, according to the constraint relation from Eq.~\eqref{Omegade_constraint}, we have \( \Omega_{d0} = 1 - \Omega_{m0} \). The expressions in eq.~\eqref{H_MCJG} and eq.~\eqref{H_MCAG} bridge the observational datasets of Hubble parameters with the theoretical frameworks of MCJG and MCAG models.

\subsection{Cosmic Chronometers (CC) and Baryon Acoustic Oscillations (BAO) Dataset}
For this work, we utilized a combined dataset that includes 26 Baryon Acoustic Oscillations (BAO) measurements and 31 Cosmic Chronometers (CC) data points, giving a total of 57 observational constraints. Cosmic chronometers provide a direct way to measure the Hubble parameter, $H(z)$, across different redshifts. This method relies on estimating the age differences between passive galaxies, which are dominated by older stellar populations and have stopped forming stars. One of the key advantages of the CC dataset is its independence from any specific cosmological model, making it a robust tool to probe the expansion history of the universe. The CC dataset consists of 31 $H(z)$ measurements across the redshift range $0.07 \leq z \leq 1.97$, collected from various studies in the literature \cite{stern2010cosmic, moresco2012new, moresco2012improved, zhang2014four, moresco2015raising, moresco20166}.
\\BAO data provides a standard ruler for measuring the angular diameter distance and the Hubble parameter at different redshifts, based on the imprint of sound waves in the early universe. The BAO dataset employed in this study consists of 26 measurements, including isotropic and anisotropic data from galaxy surveys such as SDSS \cite{ross2015clustering, alam2017clustering, gil2020completed, raichoor2021completed, hou2021completed, des2020completed}, DES \cite{abbott2022dark}, DECaLS \cite{sridhar2020clustering}, and 6dFGS \cite{beutler20116df}. These measurements cover a redshift range of $0.106 \leq z \leq 2.36$. Our analytical methodology addresses the nuanced statistical properties inherent in our cosmological datasets. The dataset comprises two distinct measurement sets: 31 uncorrelated Cosmic Chronometer observations and 26 correlated Baryon Acoustic Oscillation (BAO) measurements, each characterized by unique statistical properties. To capture the intricate statistical landscape, we implement a sophisticated likelihood estimation framework that differentiates between the independent Cosmic Chronometer measurements and the inherently correlated BAO measurements. Our approach utilizes a generalized statistical methodology encoded through the following mathematical formulation:
\begin{equation}
\chi^2_{\text{composite}} = \sum_i \sum_j \left[H_{\text{obs}}(z_i) - H_{\text{model}}(z_i)\right] , \mathbf{V}^{-1}{ij} , \left[H{\text{obs}}(z_j) - H_{\text{model}}(z_j)\right].
\end{equation}
In this formulation, $\mathbf{V}^{-1}_{ij}$ represents the inverse of the comprehensive covariance matrix, explicitly modeling the interdependencies present in BAO measurements. For the Cosmic Chronometer subset, the approach simplifies to a diagonal variance representation:
\begin{equation}
\chi^2_{\text{CC}} = \sum_i \left(\frac{H_{\text{obs}}(z_i) - H_{\text{model}}(z_i)}{\sigma_i}\right)^2 .
\end{equation}
This statistical approach ensures rigorous integration of heterogeneous cosmological observations, meticulously accounting for the distinct statistical characteristics of each measurement type. By differentially treating correlated and independent datasets, we enhance the precision and reliability of our cosmological parameter estimations.

\subsection{Pantheon+SH0ES Supernova Dataset}
We also use the Pantheon+SH0ES dataset, which combines the Pantheon sample of Type Ia supernovae (SNe Ia) with precise measurements of the Hubble constant, $H_0$, from the SH0ES project. The Pantheon+ dataset is an extended version of the original Pantheon, containing 1701 SNe Ia observations over the redshift range $0.01 \leq z \leq 2.3$, observed by ground-based telescopes and the Hubble Space Telescope (HST). The Pantheon+ dataset has been recalibrated to minimize systematic uncertainties, improving cosmological parameter estimation \cite{Scolnic_2018, Scolnic_2022}.
\\The SH0ES project provides an independent measurement of $H_0$ by directly determining the distances to SNe Ia host galaxies using Cepheid variables. The SH0ES team has reported a value of $H_0 = 73.04 \pm 1.04 \ \text{km s}^{-1} \text{Mpc}^{-1}$ \cite{Riess_2022, Riess2021}, a result that is in tension with CMB measurements and potentially hints at new physics beyond $\Lambda$CDM.
\\In this study, we employ this dataset to constrain the cosmological parameters of our models, which include \( H(z) \), the dark energy equation of state parameter \( \omega_{\text{DE}} \), and other associated quantities. The distance modulus \( \mu(z) \) is connected to the luminosity distance \( d_L(z) \) through the following relation:

\[
\mu(z) = 5 \log_{10} \left( d_L(z) \right) + 25,
\]
where $d_L(z) = (1+z) D_M(z)$, with $D_M$ for a flat universe given by:
\[
D_M(z) = \frac{c}{H_0} \int_0^z \frac{dz'}{E(z')}.
\]
Here, where \( E(z') \) is the dimensionless Hubble parameter, defined as:
\[E(z') = \frac{H(z')}{H_0}.\]
In this context, \( H(z') \) denotes the Hubble parameter at redshift \( z \), while \( H_0 \) signifies the Hubble constant. Additionally, \( c \) represents the speed of light, and \( H_0 \) is the Hubble constant at the current epoch. The log-likelihood function for the Pantheon+SH0ES dataset is expressed as:

\[
\mathcal{L}_{\text{SN}} = -\frac{1}{2} \left( \boldsymbol{\mu}_{\text{obs}} - \boldsymbol{\mu}_{\text{model}} \right)^T \mathbf{C}^{-1} \left( \boldsymbol{\mu}_{\text{obs}} - \boldsymbol{\mu}_{\text{model}} \right),
\]
where \(\boldsymbol{\mu}_{\text{obs}}\) represents the observed distance moduli, \(\boldsymbol{\mu}_{\text{model}}\) denotes the predicted distance moduli from the model, and \(\mathbf{C}^{-1}\) is the inverse of the covariance matrix. The posterior probability $\mathcal{P}(\theta)$ is proportional to the product of the likelihood and prior:
\[
\mathcal{P}(\theta) \propto \mathcal{L}_{\text{SN}} \times \pi(\theta).
\]
In this work, the absolute magnitude \( M_B \) was determined using the distance modulus \(\mu\) available in the Pantheon+SH0ES dataset. The relationship connecting the absolute magnitude \( M_B \), the distance modulus \(\mu\), and the redshift \( z \) is given by:
\begin{equation}\label{absolute_magnitude}
M_B = \mu - 5 \log_{10} \left( \frac{z (1 + z/2) c}{H_0} \right) - 25,
\end{equation}
In this expression, \(\mu\) represents the distance modulus obtained from the Pantheon+SH0ES dataset, \(z\) denotes the redshift, \(c = 3 \times 10^5 \, \mathrm{km/s}\) is the speed of light, and \(H_0 = 70 \, \mathrm{km/s/Mpc}\) is the Hubble constant used for the calculation.
The apparent magnitude data from the Pantheon+SH0ES dataset was utilized to compute \( M_B \). The low-redshift approximation for the luminosity distance was applied, assuming \( d_L(z) \approx \frac{z (1 + z/2)c}{H_0} \) for small values of \( z \). This approximation simplifies the computation and establishes a direct connection between \(\mu\) and \( M_B \). 
Theoretical predictions for \( M_B \) were consistently derived using the same conversion formula and subsequently compared with observational data. Notably, this analysis does not directly rely on the low-redshift absolute magnitude provided in the dataset. Instead, the absolute magnitude \( M_B \) was calculated using the apparent magnitude and the aforementioned formula. This approach ensures consistency in the determination of \( M_B \) for both observational and theoretical datasets, thereby enabling accurate comparisons and robust analysis.

\begin{table}[H]
\centering
\caption{Best-fit Parameters for MCJG, MCAG, and $\Lambda$CDM Models}
\renewcommand{\arraystretch}{1.5} 
\begin{tabular}{|l|c|c|c|c|c|c|}
\hline
\textbf{Model} & \textbf{Parameter} & \textbf{Pantheon+SH0ES} & \textbf{CC+BAO} \\
\hline
\multirow{7}{*}{MCJG} 
& $H_{0}$ & $68.170_{-5.526}^{+3.585}$ & $67.111_{-5.041}^{+3.584}$ \\
& $\Omega_{m0}$ & $0.262_{-0.058}^{+0.023}$ & $0.235_{-0.043}^{+0.034}$ \\
& $\omega_m$ & $0.087_{-0.061}^{+0.067}$ & $0.016_{-0.017}^{+0.035}$ \\
& $K$ & $0.353_{-0.176}^{+0.108}$ & $0.221_{-0.142}^{+0.123}$ \\
& $\alpha$ & $-0.383_{-0.181}^{+0.252}$ & $-0.451_{-0.174}^{+0.220}$ \\
& $A$ & $0.334_{-0.163}^{+0.193}$ & $0.300_{-0.130}^{+0.264}$ \\
& $V_0$ & $0.933_{-0.296}^{+0.495}$ & $1.064_{-0.369}^{+0.406}$ \\
\hline
\multirow{8}{*}{MCAG} 
& $H_{0}$ & $72.045_{-2.659}^{+0.748}$ & $68.842_{-3.520}^{+1.572}$ \\
& $\Omega_{m0}$ & $0.286_{-0.020}^{+0.011}$ & $0.266_{-0.024}^{+0.024}$ \\
& $\omega_m$ & $0.152_{-0.045}^{+0.055}$ & $0.011_{-0.028}^{+0.029}$ \\
& $e$ & $0.071_{-0.043}^{+0.097}$ & $0.088_{-0.059}^{+0.076}$ \\
& $c$ & $0.346_{-0.135}^{+0.125}$ & $0.376_{-0.145}^{+0.080}$ \\
& $\alpha$ & $0.576_{-0.372}^{+0.301}$ & $0.483_{-0.332}^{+0.282}$ \\
& $A$ & $1.386_{-0.618}^{+0.310}$ & $1.184_{-0.420}^{+0.576}$ \\
& $V_0$ & $14.959_{-1.986}^{+2.186}$ & $15.039_{-1.801}^{+2.009}$ \\
\hline
\multirow{2}{*}{$\Lambda$CDM} 
& $H_{0}$ & $73.079_{-0.158}^{+0.157}$ & $69.989_{-1.197}^{+0.876}$ \\
& $\Omega_{m0}$ & $0.341_{-0.014}^{+0.006}$ & $0.268_{-0.015}^{+0.017}$ \\
\hline
\end{tabular}
\label{tab:MCJG_MCAG_best_fit_params}
\end{table}

\subsection{Information Criteria}
The evaluation framework for model assessment incorporates three statistical metrics: the Akaike Information Criterion (AIC), Bayesian Information Criterion (BIC), and Deviance Information Criterion (DIC). These metrics provide complementary perspectives on model performance.
\\The mathematical formulation of the AIC encompasses both model complexity and goodness of fit:
\[
\text{AIC} = -2 \ln(L_{\text{max}}) + 2k + \frac{2k(k+1)}{N_{\text{tot}} - k - 1}
\quad.\]
In this expression, $L_{\text{max}}$ represents the maximum likelihood value, $k$ quantifies the number of model parameters, and $N_{\text{tot}}$ indicates the total quantity of observational data-points.
The BIC expression incorporates a complexity penalty that scales with sample size:
\[
\text{BIC} = -2 \ln(L_{\text{max}}) + k \ln(N_{\text{tot}})\quad.
\]
where the variables maintain their previous definitions: $L_{\text{max}}$ for maximum likelihood, $k$ for parameter count, and $N_{\text{tot}}$ for the observational dataset size.
The DIC computation involves multiple components and is expressed as:
\[
\text{DIC} = \bar{D} + p_D \quad.
\]
The mean deviance $\bar{D}$ across posterior samples is calculated through:
\[
\bar{D} = \frac{1}{S} \sum_{i=1}^S D(\theta_i)\quad,
\]
where $S$ represents the total posterior sample count and $D(\theta_i)$ denotes individual sample deviance. The effective parameter count $p_D$ is determined by:
\[
p_D = \bar{D} - D(\hat{\theta})\quad,
\]
with $D(\hat{\theta})$ representing the deviance evaluated at the posterior parameter means.
To compare the different models, we compute the AIC, BIC, and DIC for each model and for the $\Lambda$CDM model. The results are presented in Table~\ref{tab:MCJG_MCAG_model_metrics}. For comparison purposes, we calculate the relative difference in the Information Criterion (IC) values, defined as:
\[
\Delta IC_{\text{model}} = IC_{\text{model}} - IC_{\text{min}}\quad,
\]
where \( IC_{\text{min}} \) is the smallest IC value among the models under consideration \cite{anagnostopoulos2020observational}. According to the Jeffreys scale \cite{jeffreys1998theory}, the interpretation of the \( \Delta IC \) values is as follows: a \( \Delta IC \leq 2 \) indicates that the model is statistically indistinguishable from the reference model, a \( \Delta IC \) between 2 and 6 suggests moderate evidence against the model, and a \( \Delta IC \geq 10 \) indicates strong evidence against the model.

\begin{table}[H]
\centering
\caption{Information Criterion and $\Delta$IC values for different models}
\begin{tabular}{|l|c|c|c|c|c|c|c|}
\hline
\textbf{Model} & \textbf{Criteria} & \textbf{AIC} & \textbf{BIC} & \textbf{DIC} & $\Delta$ \textbf{AIC} & $\Delta$ \textbf{BIC} & $\Delta$ \textbf{DIC} \\
\hline
\multicolumn{8}{|c|}{\textbf{Pantheon+SH0ES Dataset}} \\
\hline
MCJG & Value & 1764.20 & 1775.27 & 1714.28 & 6.47 & 6.66 & 0.00 \\
MCAG & Value & 1765.83 & 1773.34 & 1721.92 & 8.10 & 4.73 & 7.64 \\
$\Lambda$CDM & Value & 1757.73 & 1768.61 & 1718.97 & 0.00 & 0.00 & 4.69 \\
\hline
\multicolumn{8}{|c|}{\textbf{CC+BAO Dataset}} \\
\hline
MCJG & Value & 42.44 & 47.75 & 33.25 & 6.20 & 7.42 & 0.00 \\
MCAG & Value & 41.20 & 45.54 & 35.78 & 5.96 & 5.21 & 2.53 \\
$\Lambda$CDM & Value & 36.24 & 40.33 & 36.22 & 0.00 & 0.00 & 2.97 \\
\hline
\end{tabular}
\label{tab:MCJG_MCAG_model_metrics}
\end{table}

\subsection{Results and Findings}
\begin{figure}[htbp]
    \centering
    \begin{subfigure}[b]{0.75\textwidth}
        \includegraphics[width=\textwidth]{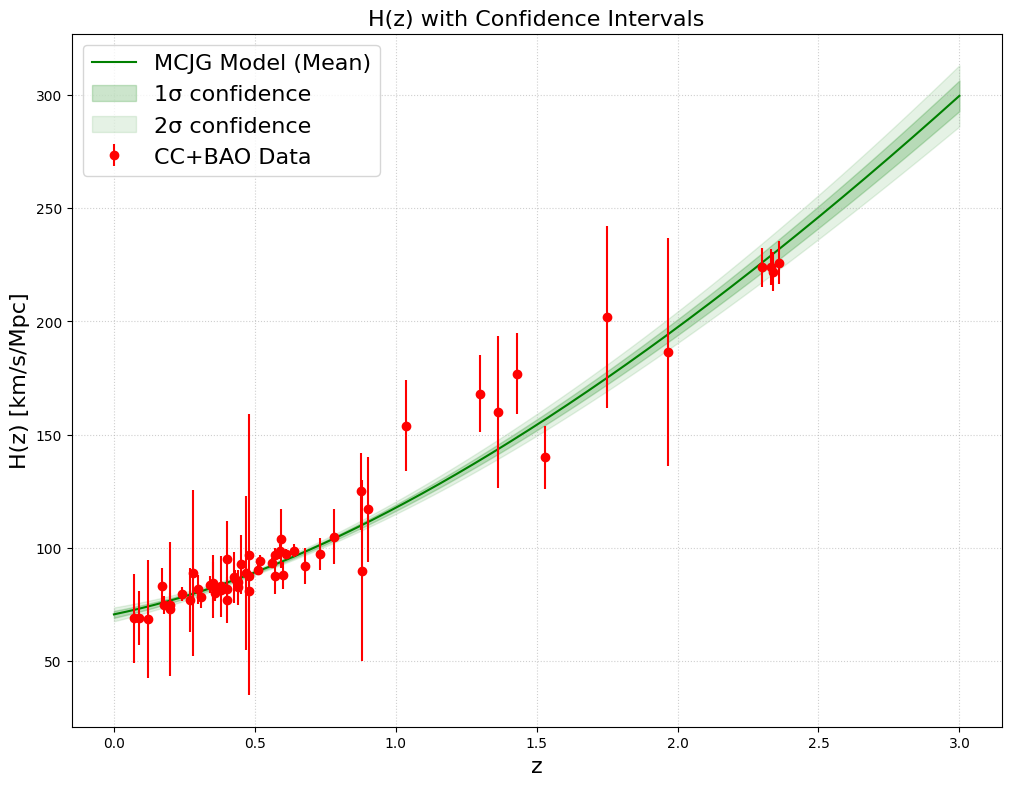}
        \caption{}
        \label{fig:Hubble_MCJG}
    \end{subfigure}
    \begin{subfigure}[b]{0.75\textwidth}
        \includegraphics[width=\textwidth]{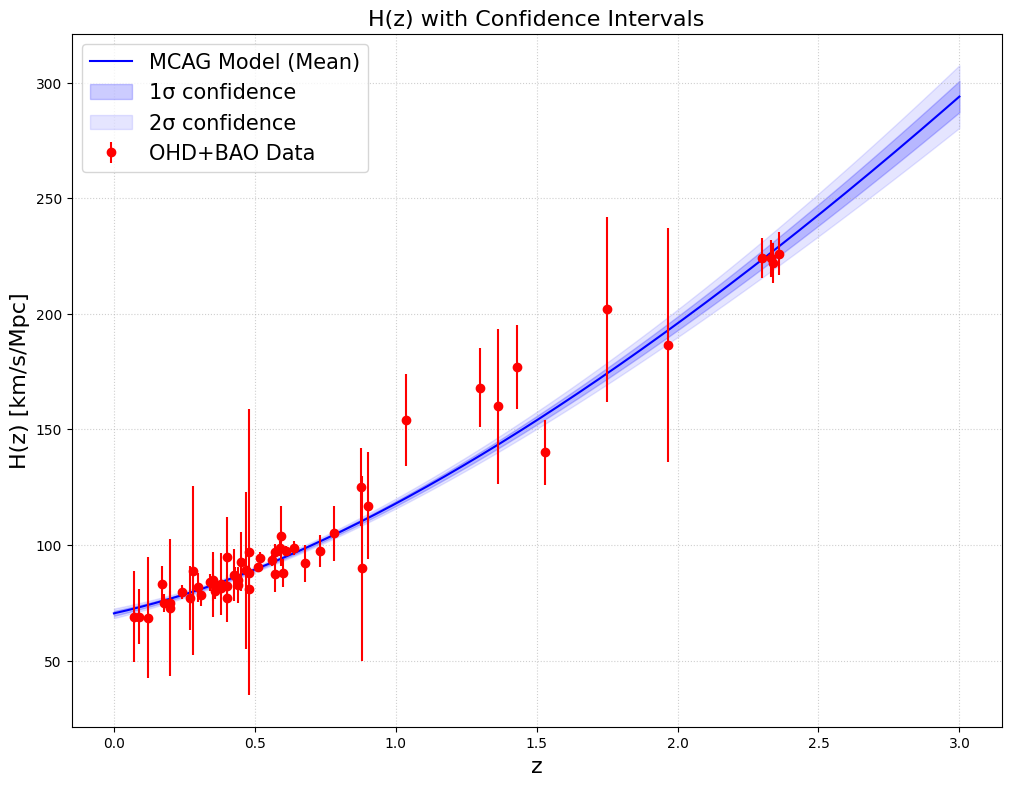}
        \caption{}
        \label{fig:Hubble_MCAG}
    \end{subfigure}
    \caption{Plot of Hubble parameter at 1$\sigma$ and 2$\sigma$ confidence level for the fig: (a) MCJG model and fig: (b) MCAG model using 57 CC + BAO data points.}
    \label{fig:Hubble_MCJG_MCAG}
\end{figure}

\begin{figure}[htbp]
    \centering
    \begin{subfigure}[b]{0.75\textwidth}
        \includegraphics[width=\textwidth]{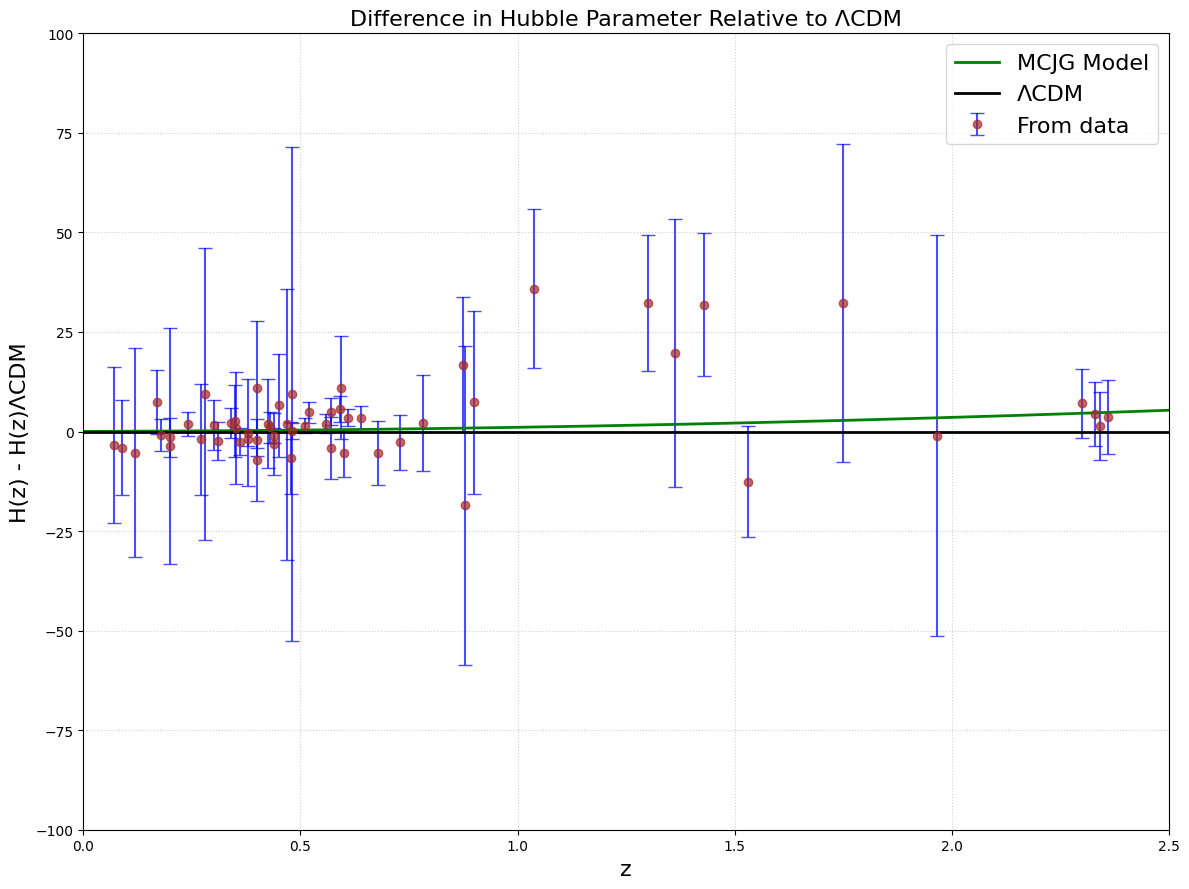}
        \caption{}
        \label{fig:Relative Hubble_MCJG}
    \end{subfigure}
    \begin{subfigure}[b]{0.75\textwidth}
        \includegraphics[width=\textwidth]{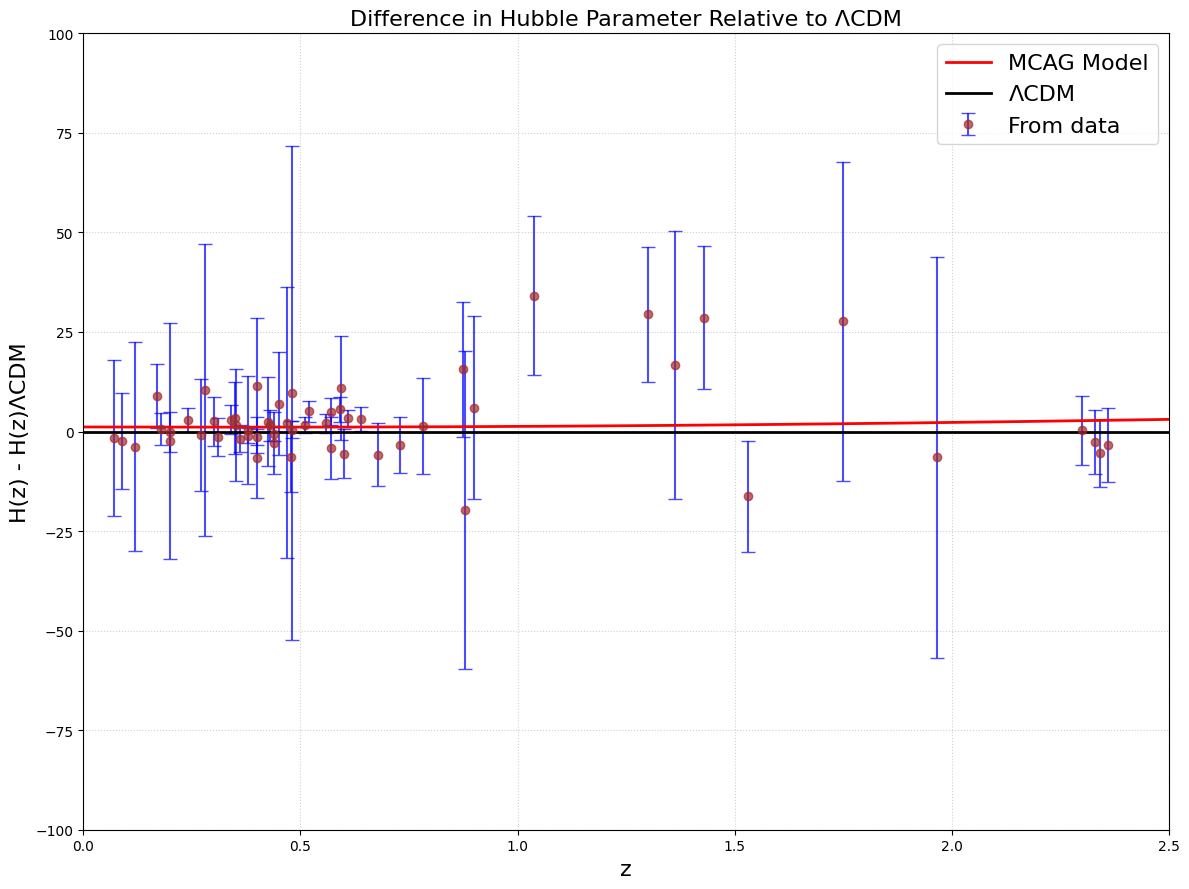}
        \caption{}
        \label{fig:Relative Hubble_MCAG}
    \end{subfigure}
    \caption{Comparison of variation of difference in Hubble Parameter between $\Lambda$CDM and the fig: (a) MCJG model and fig: (b) MCAG model gravity using 57 CC + BAO data points.}
    \label{fig:Relative_Hubble_MCJG_MCAG}
\end{figure}

\begin{figure}[htbp]
    \centering
    \begin{subfigure}[b]{0.75\textwidth}
        \includegraphics[width=\textwidth]{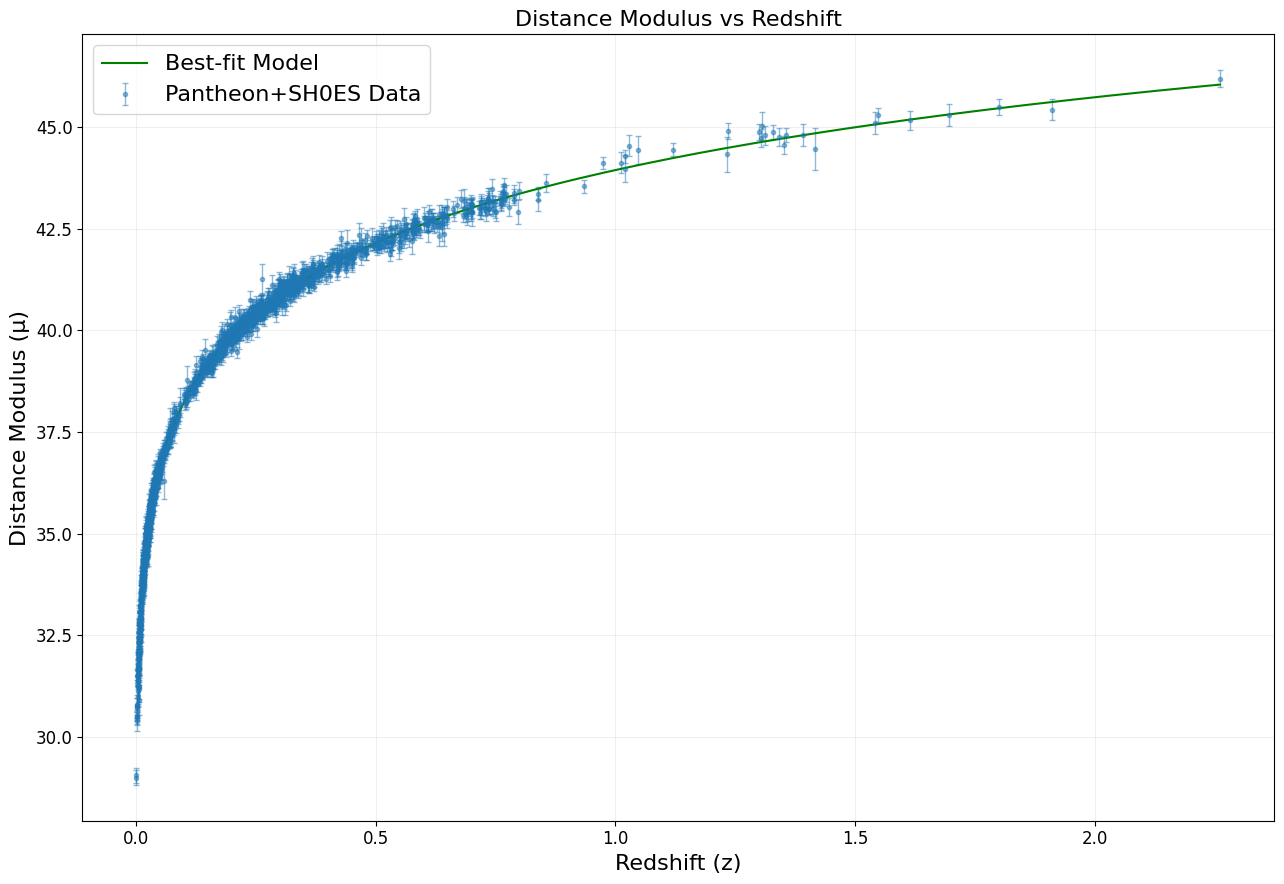}
        \caption{}
        \label{fig:Distance_modulus_MCJG}
    \end{subfigure}
    \begin{subfigure}[b]{0.75\textwidth}
        \includegraphics[width=\textwidth]{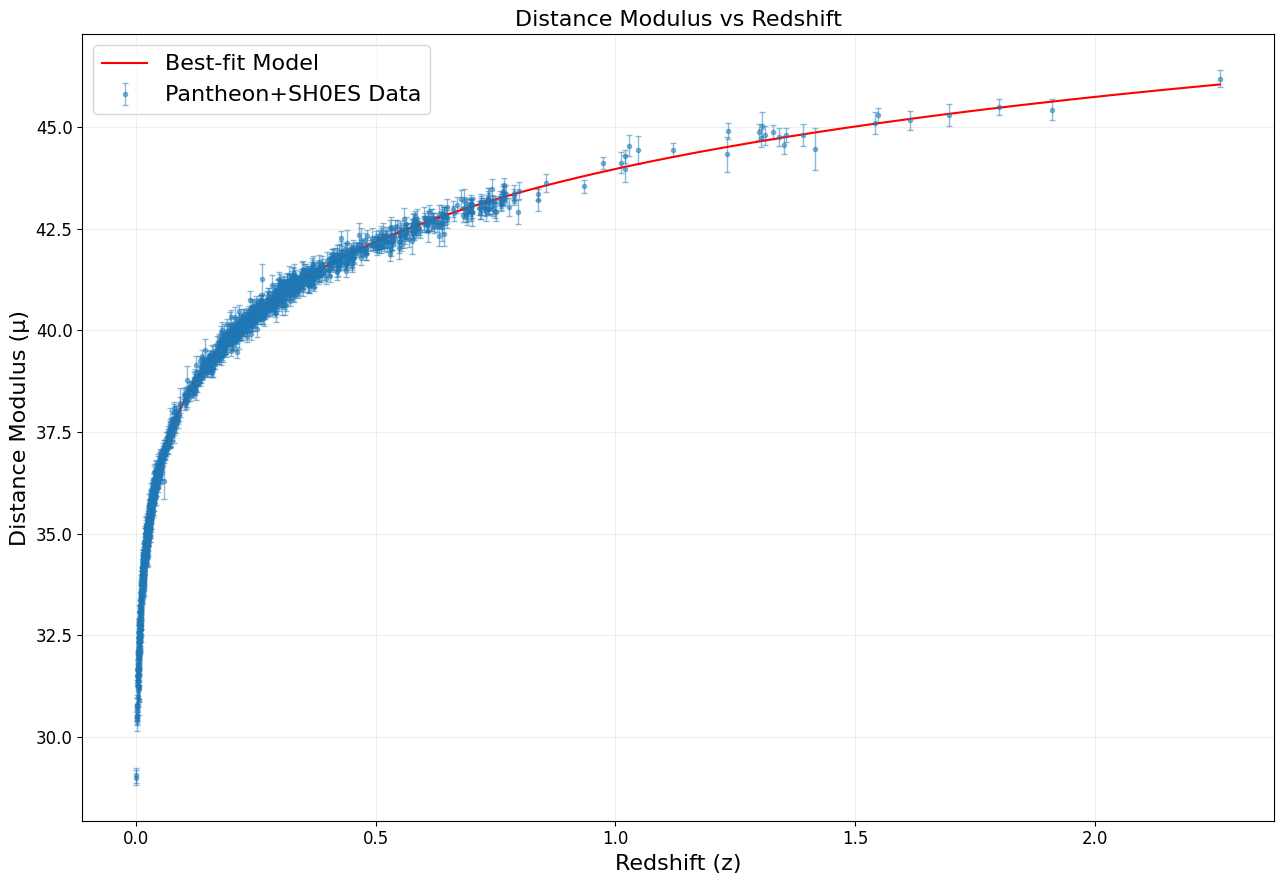}
        \caption{}
        \label{fig:Distance_modulus_MCAG}
    \end{subfigure}
    \caption{Best fit plot of Distance modulus against redshift for fig: (a) MCJG model and fig: (b) MCAG model gravity using 1701 Pantheon+SH0ES data points.}
    \label{fig:Distance_modulus_MCJG_MCAG}
\end{figure}

\begin{figure}[htbp]
    \centering
    \begin{subfigure}[b]{0.75\textwidth}
        \includegraphics[width=\textwidth]{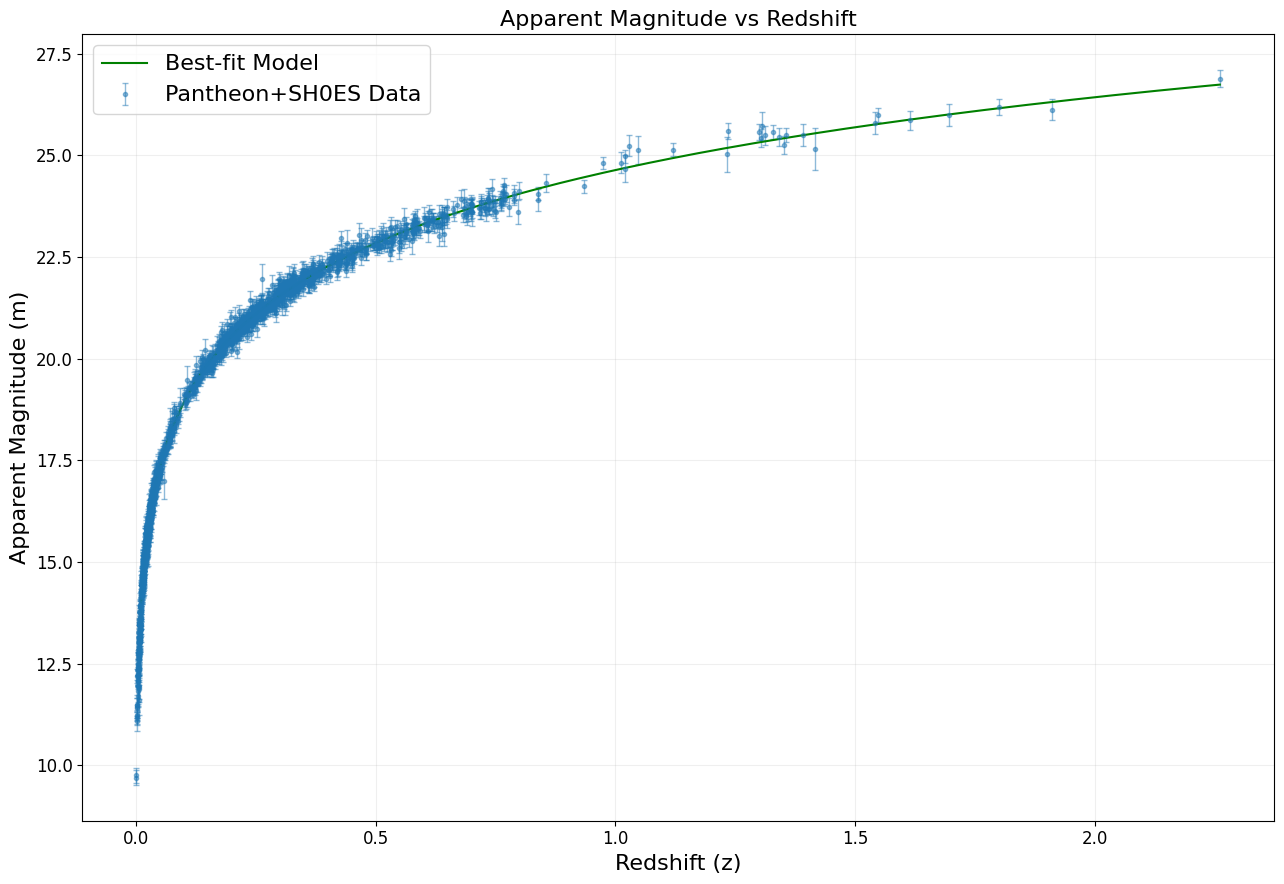}
        \caption{}
        \label{fig:Apparent_magnitude_MCJG}
    \end{subfigure}
    \begin{subfigure}[b]{0.75\textwidth}
        \includegraphics[width=\textwidth]{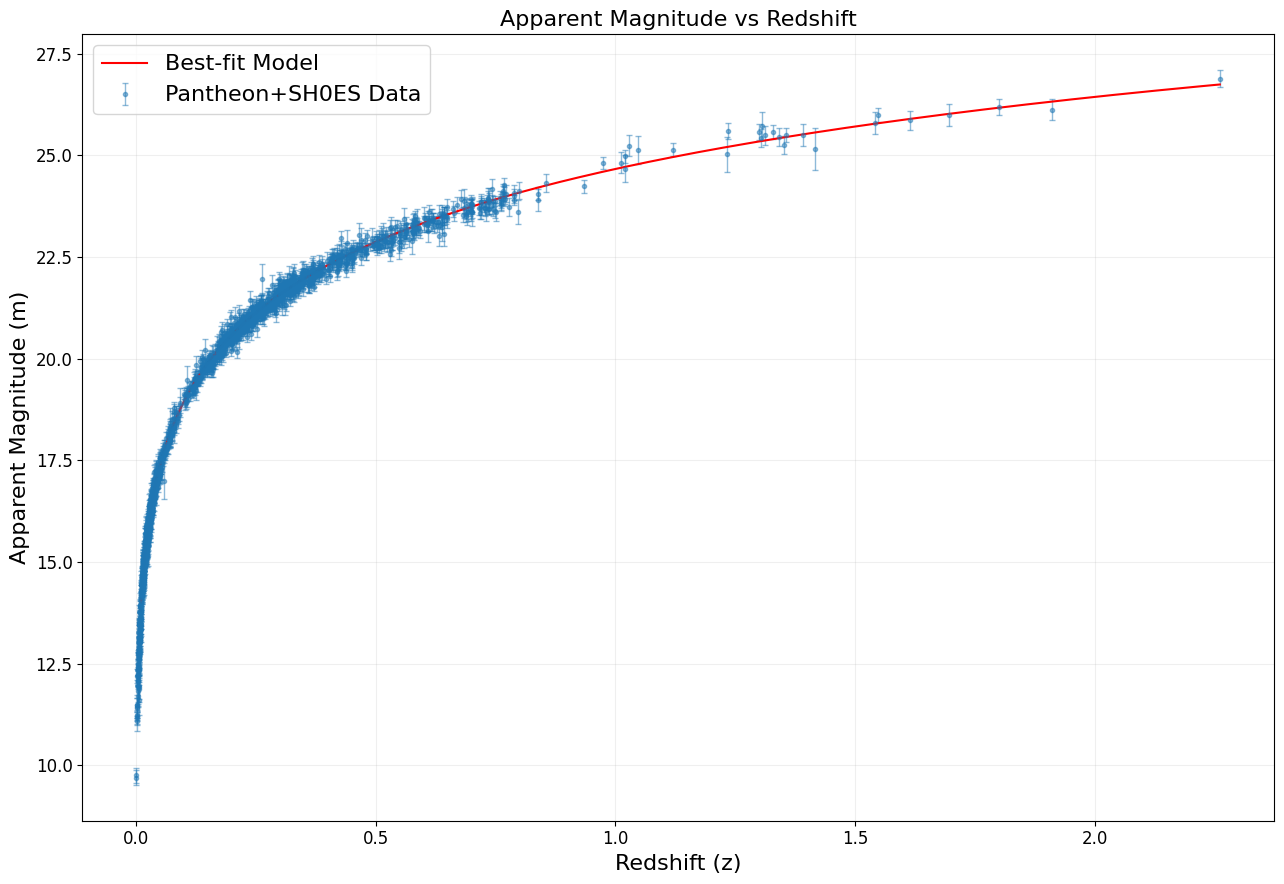}
        \caption{}
        \label{fig:Apparent_magnitude_MCAG}
    \end{subfigure}
    \caption{Best fit plot of Apparent magnitude against redshift for fig: (a) MCJG model and fig: (b) MCAG model gravity using 1701 Pantheon+SH0ES data points.}
    \label{fig:Apparent_magnitude_MCJG_MCAG}
\end{figure}

\begin{figure}[htbp]
    \centering
    \begin{subfigure}[b]{0.75\textwidth}
        \includegraphics[width=\textwidth]{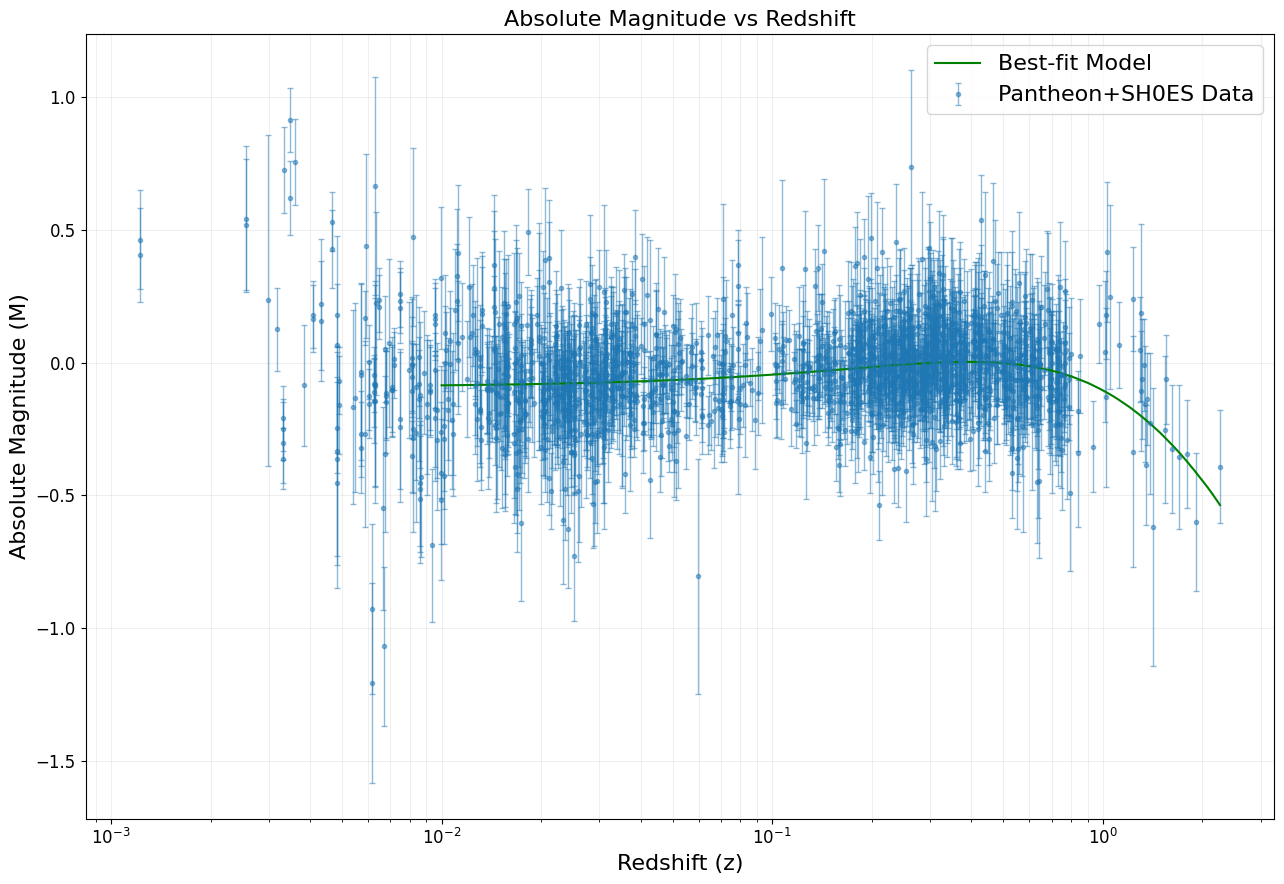}
        \caption{}
        \label{fig:Absolute_magnitude_MCJG}
    \end{subfigure}
    \begin{subfigure}[b]{0.75\textwidth}
        \includegraphics[width=\textwidth]{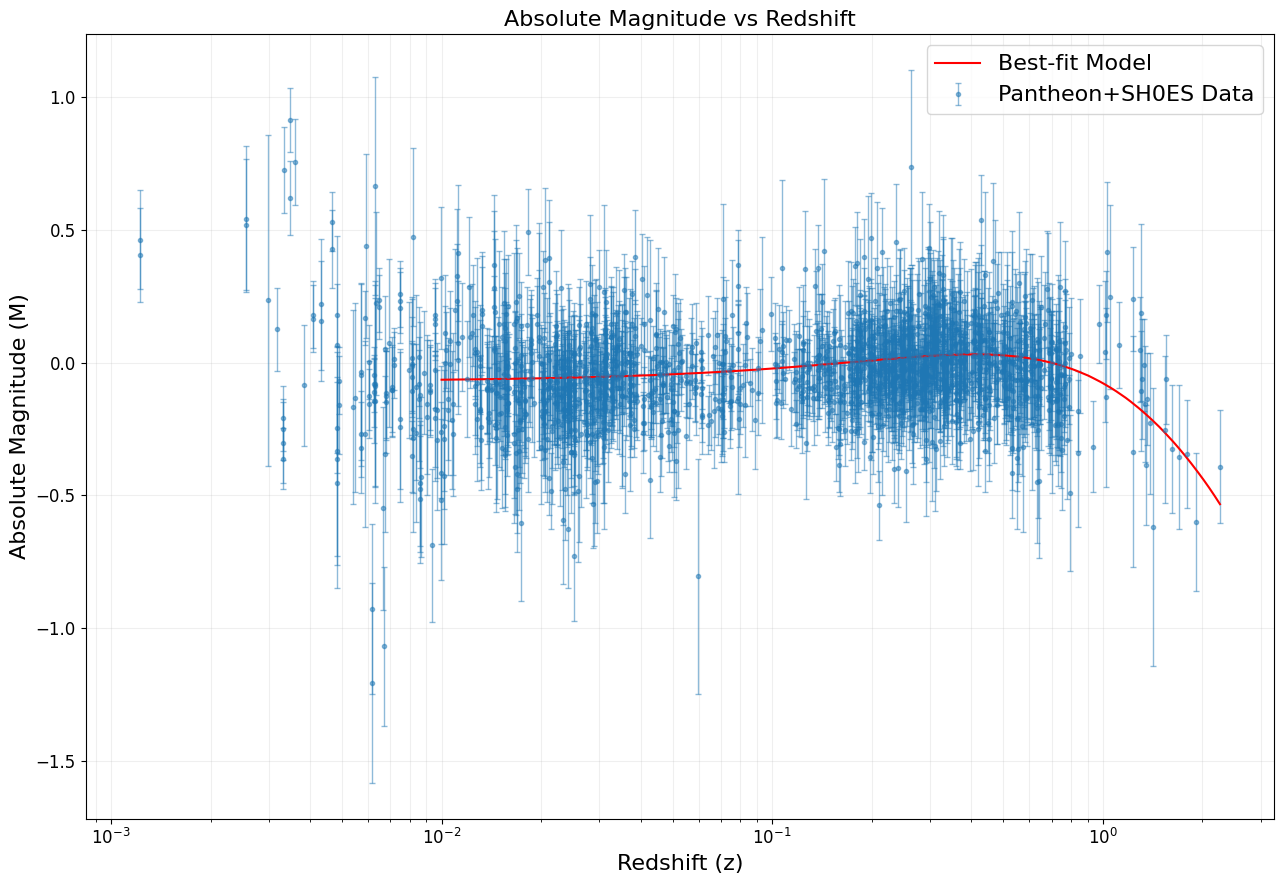}
        \caption{}
        \label{fig:Absolute_magnitude_MCAG}
    \end{subfigure}
    \caption{Best fit plot of Absolute magnitude against redshift for fig: (a) MCJG model and fig: (b) MCAG model gravity using 1701 Pantheon+SH0ES data points.}
    \label{fig:Absolute_magnitude_MCJG_MCAG}
\end{figure}

We evaluate the effectiveness of our models by comparing them against key cosmological data such as the Pantheon+SH0ES dataset. The Hubble constant values for the MCJG and MCAG models are as follows: for MCJG, $H_0 = 68.170^{+3.585}_{-5.526}$ km s$^{-1}$ Mpc$^{-1}$ (Pantheon+SH0ES) and $H_0 = 67.111^{+3.584}_{-5.041}$ km s$^{-1}$ Mpc$^{-1}$ (CC+BAO); for MCAG, $H_0 = 72.045^{+0.748}_{-2.659}$ km s$^{-1}$ Mpc$^{-1}$ (Pantheon+SH0ES) and $H_0 = 68.842^{+1.572}_{-3.520}$ km s$^{-1}$ Mpc$^{-1}$ (CC+BAO).
The analysis of the Hubble constant ($H_0$) values for both MCJG and MCAG models reveals interesting insights into the Hubble tension problem. For MCJG, we observe $H_0 = 68.170^{+3.585}_{-5.526}$ km s$^{-1}$ Mpc$^{-1}$ (Pantheon+SH0ES), which aligns closely with the Planck collaboration's CMB-based measurements\cite{refId0}. This suggests that MCJG maintains consistency with early-universe observations. However, MCAG yields $H_0 = 72.045^{+0.748}_{-2.659}$ km s$^{-1}$ Mpc$^{-1}$ (Pantheon+SH0ES), showing better agreement with local measurements from SH0ES collaboration ($73.24\pm 1.74$ km s$^{-1}$ Mpc$^{-1}$)\cite{riess2019large}. This divergence between the two models can be physically interpreted as follows:

\begin{itemize}
    \item \textbf{MCJG Behavior:} The closer alignment with Planck data suggests that MCJG better preserves the standard cosmological evolution from the early universe. This could indicate that the model maintains stronger coupling between matter and radiation components throughout cosmic evolution, leading to a more traditional expansion history.
    
    \item \textbf{MCAG Behavior:} The higher $H_0$ value and consistency with SH0ES measurements suggests that MCAG potentially incorporates mechanisms that allow for additional expansion in the late universe, possibly through modified gravitational dynamics or altered cosmic acceleration phases.
\end{itemize}
Regarding the Hubble tension, these models present distinct physical mechanisms for addressing the discrepancy:

\begin{enumerate}
    \item \textbf{MCJG's Approach:} This model favors the early-universe measurements with $H_0 = 68.170^{+3.585}_{-5.526}$ km s$^{-1}$ Mpc$^{-1}$, aligning with Planck observations. The larger uncertainties indicate that the model accommodates potential variations in the cosmic expansion rate through:
    \begin{itemize}
        \item Possible modifications to the standard matter-radiation coupling in the early universe.
        \item Flexibility in the dark energy equation of state, allowing for dynamical evolution without violating the cosmic expansion history constraints.
        \item Maintenance of consistency with BAO measurements ($H_0 = 67.111^{+3.584}_{-5.041}$ km s$^{-1}$ Mpc$^{-1}$), suggesting compatibility with large-scale structure formation.
    \end{itemize}
    
    \item \textbf{MCAG's Approach:} With $H_0 = 72.045^{+0.748}_{-2.659}$ km s$^{-1}$ Mpc$^{-1}$, this model provides a physical resolution to the tension through:
    \begin{itemize}
        \item Modified gravitational dynamics that could affect late-time cosmic acceleration.
        \item Tighter constraints on parameter space, indicated by smaller uncertainties, suggesting a more specific mechanism for reconciling early and late-universe measurements.
        \item A natural transition between the CC+BAO value ($68.842^{+1.572}_{-3.520}$ km s$^{-1}$ Mpc$^{-1}$) and the local SH0ES measurements, indicating a physically consistent evolution of the expansion rate.
    \end{itemize}
\end{enumerate}
The CC+BAO measurements for both models ($H_0 = 67.111^{+3.584}_{-5.041}$ and $68.842^{+1.572}_{-3.520}$ km s$^{-1}$ Mpc$^{-1}$ for MCJG and MCAG respectively) provide additional support for their physical consistency, suggesting that both models maintain compatibility with large-scale structure formation while offering different approaches to the Hubble tension problem. This dichotomy between MCJG and MCAG demonstrates that the Hubble tension might be addressable through multiple theoretical frameworks, each with distinct physical implications for cosmic evolution and structure formation.
In Fig.~\ref{fig:Hubble_MCJG_MCAG}, we present the evolution of the Hubble parameter $H(z)$ against the redshift $z$ for the MCJG and MCAG models at 1$\sigma$ and 2$\sigma$ confidence level . The data points from the CC+BAO dataset are illustrated in red. Both plots in Fig.~\ref{fig:Hubble_MCJG} and Fig.~\ref{fig:Hubble_MCAG} demonstrate a well-fitted scenario of our models with the observed data. This close match with the observed data underscores the capability of our models to explain the dynamical evolution of the universe.
Fig.~\ref{fig:Relative_Hubble_MCJG_MCAG} depicts the variation in the difference between the MCJG, MCAG models and the $\Lambda$CDM model. For redshifts greater than 0.5 $( z > 0.5 )$, a noticeable deviation of both models becomes apparent when compared to cosmic chronometer (CC) and BAO measurements. This indicates that at higher redshifts, the MCJG and MCAG models deviate slightly from the predictions made by the $\Lambda$CDM model, indicating the presence of different physical processes or parameters that become more significant at these higher redshifts.
On the other hand, for redshifts less than 0.5 $( z < 0.5 )$, this deviation reduces. As the redshift decreases, the MCJG and MCAG models gradually converge towards the $\Lambda$CDM model. This observation implies that at lower redshifts, the differences between the MCJG, MCAG models, and the $\Lambda$CDM model diminish, leading to predictions that are more closely aligned with the observational data provided by the cosmic chronometers and Baryon Acoustic Oscillation.
\\Figure \ref{fig:Distance_modulus_MCJG_MCAG} presents the distance modulus ($\mu$) as a function of redshift ($z$) for both the MCJG and MCAG models, fitted to the Pantheon+SH0ES dataset. The observed data points are depicted as blue dots with error bars, while the best-fit models are represented by solid lines (green for MCJG in Fig.~\ref{fig:Distance_modulus_MCJG}, red for MCAG in Fig.~\ref{fig:Distance_modulus_MCAG}). Both models exhibit excellent agreement with the observational data across the entire redshift range ($10^{-3} < z < 2.3$). As expected in an expanding universe, the distance modulus increases monotonically with redshift. This reflects the growing distance light must travel from more distant (higher redshift) sources, leading to fainter apparent magnitudes and consequently larger distance moduli. 
At low redshifts ($z < 0.01$), both models accurately capture the near-linear relationship between distance modulus and redshift (on a log scale), a region crucial for constraining the Hubble constant ($H_0$). In the range $0.01 < z < 0.1$, a smooth transition is observed in the slope of the curve, indicating the increasing influence of cosmic expansion on the distance-redshift relation. The close alignment between both models and the observational data suggests that the MCJG and MCAG frameworks effectively describe the expansion history of the universe, as observed through Type Ia supernovae.
\\The analysis of the Apparent Magnitude vs Redshift plots in Fig.~\ref{fig:Apparent_magnitude_MCJG_MCAG} for the Modified Chaplygin-Jacobi Gas and the Modified Chaplygin-Abel Gas models reveals remarkable similarities in their overall trends, with both showing an increase in apparent magnitude with increasing redshift. This behavior aligns with cosmological expectations, where more distant objects appear fainter. Both models demonstrate strong compatibility with the Pantheon+SH0ES observational data, particularly at higher redshifts $(z > 0.1)$. However, subtle differences emerge in the low (z < 0.01) and intermediate $(0.01 < z < 0.1)$ redshift regimes. The MCAG model exhibits a marginally better fit for the nearest objects, while the MCJG model appears to provide a slightly closer fit in the intermediate range. These nuanced distinctions suggest that both models effectively capture the essential features of cosmic acceleration and the transformation from matter-dominated to dark energy-dominated epochs.
The close alignment between both models and the observational data underscores their viability as candidates for describing the universe's expansion history. The subtle variations observed, especially in the low and intermediate redshift ranges, carry significant implications for our understanding of recent cosmic expansion and the nature of dark energy. These differences may offer valuable insights into constraining the current value of the Hubble parameter ($H_0$) and refining our treatment of the transition from matter domination to dark energy domination. While visual inspection reveals these subtle distinctions, a more rigorous statistical analysis would be necessary to quantitatively discriminate between the models.
\\The Absolute Magnitude vs Redshift plots for the Modified Chaplygin-Jacobi Gas in Fig.~\ref{fig:Absolute_magnitude_MCJG} and the Modified Chaplygin-Abel Gas in Fig.~\ref{fig:Absolute_magnitude_MCAG} reveal important insights into the cosmic distance-redshift relationship and the expansion history of the universe. Both plots display the Pantheon+SH0ES data as blue points with error bars, showing a wide spread of absolute magnitudes across the redshift range. The best-fit models, represented by green (MCJG) and red (MCAG) lines respectively, demonstrate subtle but significant differences in their predictions. At low redshifts ($z < 0.01$), both models show a slight upward trend in absolute magnitude. In the intermediate redshift range ($0.01 < z < 0.1$), the models predict a nearly constant absolute magnitude, with the MCJG model showing a slightly flatter trend. At high redshifts (z > 0.1), both models predict a gradual decrease in absolute magnitude, with the MCAG model exhibiting a steeper decline beyond $z \approx 1$.
The physical significance of these Absolute Magnitude vs Redshift plots is profound, offering crucial insights into the nature of cosmic expansion and the properties of dark energy. The absolute magnitude of Type Ia supernovae, serving as standard candles, should remain constant in a static universe. However, the observed variations with redshift directly reflect the dynamic nature of cosmic expansion. The subtle differences between the MCJG and MCAG models, particularly at high redshifts, suggest slightly different predictions for the evolution of dark energy. The MCAG model's steeper decline in absolute magnitude at high redshifts could indicate a more rapid change in the dark energy equation of state, potentially pointing to a dynamical dark energy scenario. Conversely, the MCJG model's flatter trend might suggest a dark energy behavior closer to a cosmological constant. These distinctions are crucial for getting a profound insight into the nature of dark energy and its role in cosmic acceleration, potentially offering clues to discriminate between various dark energy models and modified gravity theories.
\\Figure~\ref{fig:Omega_total_MCJG_MCAG} depicts the evolution of the normalized energy density parameters $\Omega(z)$ as a function of redshift $z$ for the MCJG and MCAG cosmological models. It emphasizes two key components: the matter density parameter \( \Omega_m(z) \) and the dark energy density parameter \( \Omega_{de}(z) \). A vertical blue dashed line at $z=0$ indicates the present critical density. As we move towards higher redshifts, representing earlier times in the universe's history, the matter density parameter increases, indicating that the universe was primarily dominated by matter in the past. In contrast, the dark energy density parameter decreases with increasing redshift, implying that dark energy played a less prominent role in the earlier universe but has become more dominant in the present era.
\begin{figure}[htbp]
    \centering
    \begin{subfigure}[b]{0.75\textwidth}
        \includegraphics[width=\textwidth]{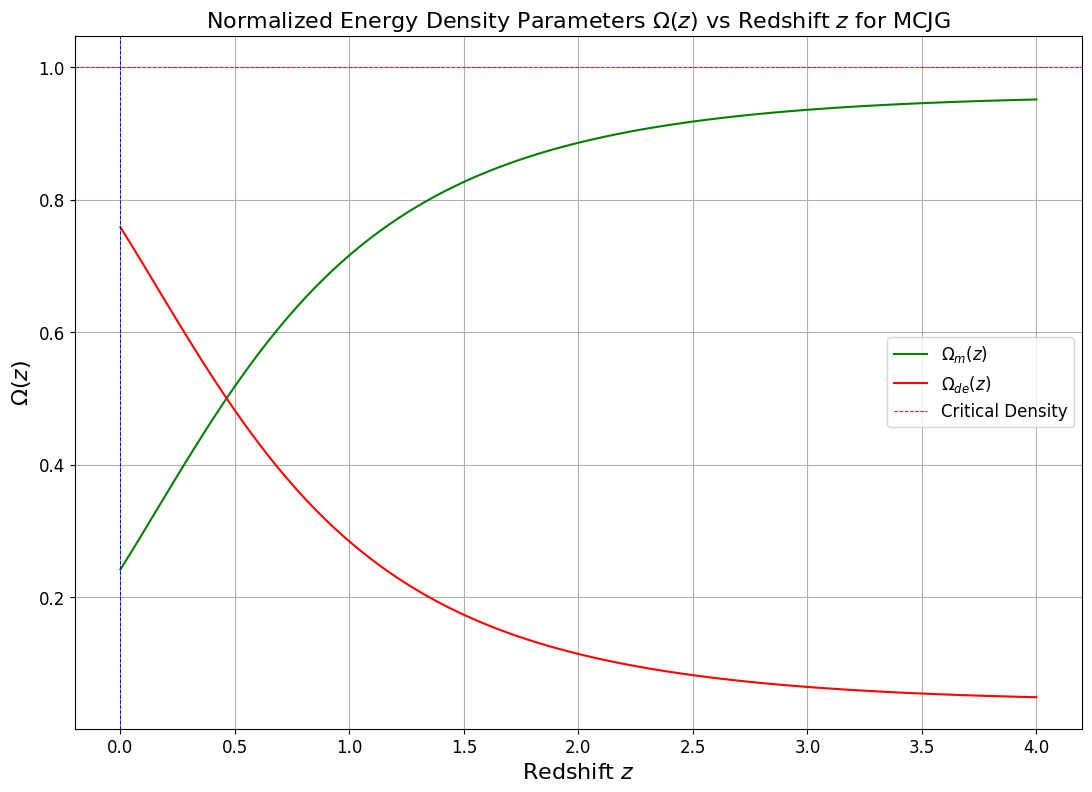}
        \caption{}
        \label{fig:Omega_total_MCJG}
    \end{subfigure}
    \begin{subfigure}[b]{0.75\textwidth}
        \includegraphics[width=\textwidth]{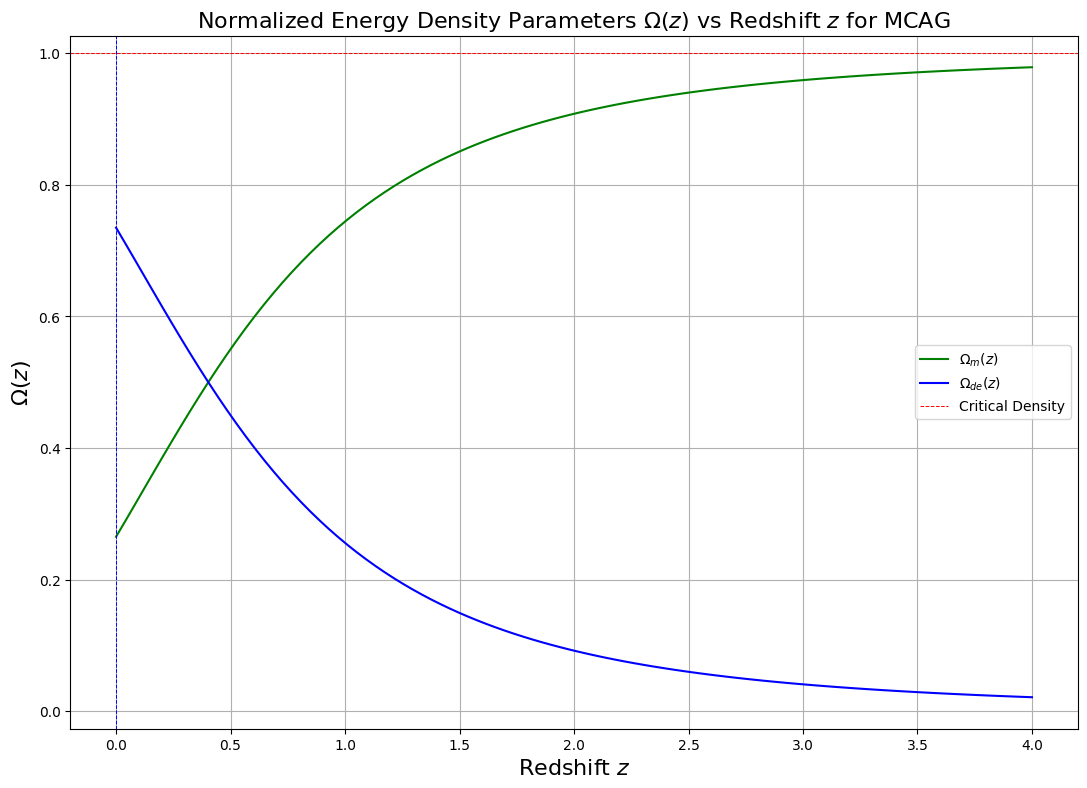}
        \caption{}
        \label{fig:Omega_total_MCAG}
    \end{subfigure}
    \caption{Plot of density parameter components of matter and dark energy for the fig: (a) MCJG model and fig: (b) MCAG model.}
    \label{fig:Omega_total_MCJG_MCAG}
\end{figure}
The two curves intersect around $z \approx 0.5$, marking the transition from a matter-dominated universe to one where dark energy governs the expansion. Importantly, the sum of $\Omega_m$ and $\Omega_{de}$ equals 1 throughout, consistent with the properties of a flat universe.
\\From Table~\ref{tab:MCJG_MCAG_best_fit_params}, we observe several key differences and similarities in the parameter values between the MCJG and MCAG models for both the Pantheon+SH0ES and CC+BAO datasets. For the matter density parameter at the current epoch, $\Omega_{m0}$, the MCJG model estimates it as $0.262_{-0.058}^{+0.023}$ from Pantheon+SH0ES and $0.235_{-0.043}^{+0.034}$ from CC+BAO. In contrast, the MCAG model shows slightly higher values, with $\Omega_{m0} = 0.286_{-0.020}^{+0.011}$ (Pantheon+SH0ES) and $0.266_{-0.024}^{+0.024}$ (CC+BAO). These differences suggest that the MCAG model generally assumes a higher matter density at the present epoch compared to the MCJG model, but both are within reasonable ranges compared to standard cosmological models and observational data\cite{aghanim2020planck}.
The equation of state parameter for matter, $\omega_m$, reflects similar behavior in both models. For the MCJG model, $\omega_m = 0.087_{-0.061}^{+0.067}$ from Pantheon+SH0ES and $0.016_{-0.017}^{+0.035}$ from CC+BAO. The MCAG model exhibits a slightly larger $\omega_m$ for Pantheon+SH0ES, $\omega_m = 0.152_{-0.045}^{+0.055}$, while the CC+BAO value is close to zero at $\omega_m = 0.011_{-0.028}^{+0.029}$. These results indicate that both models generally assume a nearly dust-like behavior for matter, though the MCAG model suggests a higher equation of state parameter for matter than the MCJG model, especially when using the Pantheon+SH0ES dataset.
The $K$ parameter for the MCJG model is constrained as $K = 0.353_{-0.176}^{+0.108}$ (Pantheon+SH0ES) and $K = 0.221_{-0.142}^{+0.123}$ (CC+BAO). This parameter does not have an equivalent in the MCAG model, but the slightly higher values from Pantheon+SH0ES suggest that this parameter plays a more prominent role in the MCJG model, particularly in scenarios involving higher redshift data. This value is in close agreement with the previous observational analysis done be Debnath~\cite{debnath_MCAG_MCJG}.
In the MCAG model, the parameter $e$ is constrained as $e = 0.071_{-0.043}^{+0.097}$ from Pantheon+SH0ES and $e = 0.088_{-0.059}^{+0.076}$ from CC+BAO. These relatively low values suggest that $e$ has a smaller impact on the overall evolution of the universe in the MCAG model compared to $K$ in the MCJG model. The parameter $c$ in the MCAG model is constrained as $c = 0.346_{-0.135}^{+0.125}$ from Pantheon+SH0ES and $c = 0.376_{-0.145}^{+0.080}$ from CC+BAO. These values show consistency across datasets, indicating that $c$ plays a comparable role in both Pantheon+SH0ES and CC+BAO observations. Both the constrained values of $e$ and $c$ are in alignment with \cite{debnath_MCAG_MCJG}.
Both models include the parameter $\alpha$, though the signs of $\alpha$ differ. For the MCJG model, $\alpha = -0.383_{-0.181}^{+0.252}$ (Pantheon+SH0ES) and $\alpha = -0.451_{-0.174}^{+0.220}$ (CC+BAO), while for the MCAG model, $\alpha = 0.576_{-0.372}^{+0.301}$ (Pantheon+SH0ES) and $\alpha = 0.483_{-0.332}^{+0.282}$ (CC+BAO). The positive values of $\alpha$ in the MCAG model contrast with the negative values in the MCJG model, suggesting different dynamics in the models regarding the influence of this parameter. The small values of $\alpha$ are in close agreement with the previous works on Chaplygin gas~\cite{debnath_MCAG_MCJG,fabris2010matter}. The negative value of $\alpha$ for the MCJG shows its resemblance with the Generalised Chaplygin gas (GCG) model where a increasingly small and negative value of $\alpha$ is observationally preferred as seen in ref.~\cite{lu2009observational, fabris2010matter}.
For the parameter $A$ the MCJG model gives $A = 0.334_{-0.163}^{+0.193}$ (Pantheon+SH0ES) and $A = 0.300_{-0.130}^{+0.264}$ (CC+BAO). The MCAG model exhibits higher values, with $A = 1.386_{-0.618}^{+0.310}$ (Pantheon+SH0ES) and $A = 1.184_{-0.420}^{+0.576}$ (CC+BAO), indicating a more significant influence of the potential energy term in the MCAG model compared to the MCJG model. Though the value of parameter $A$ is in close agreement with \cite{debnath_MCAG_MCJG} for MCJG model but it completely differs in the case of MCAG.
Finally, the parameter $V_0$, which is the volume parameter, differs substantially between the two models. For the MCJG model, $V_0 = 0.933_{-0.296}^{+0.495}$ (Pantheon+SH0ES) and $V_0 = 1.064_{-0.369}^{+0.406}$ (CC+BAO). In the MCAG model, however, $V_0$ is much larger, with $V_0 = 14.959_{-1.986}^{+2.186}$ (Pantheon+SH0ES) and $V_0 = 15.039_{-1.801}^{+2.009}$ (CC+BAO). This significant difference suggests that the MCAG model assumes a much stronger contribution from the dark energy potential than the MCJG model, which may lead to different cosmic evolution predictions in each scenario.
Lastly, the age of the universe, $t_{\text{age}}$, is found to be $13.78^{+0.14}_{-0.17}$ Gyr for the MCJG model and $13.93^{+0.16}_{-0.16}$ Gyr for the MCAG model. These values show consistency across datasets and align well with the current estimates of the universe's age from observational data, which support an age of approximately $13.8$ Gyr \cite{aghanim2020planck}.

\section{Conclusion}\label{Sect:conclusion}
Our investigation presents a comprehensive analysis of the thermodynamic characteristics exhibited by two cosmological models: the Modified Chaplygin-Jacobi Gas and Modified Chaplygin-Abel Gas in the context of universal evolution. Through our analysis, we have derived a novel thermal equation of state expressing pressure as a function of temperature and volume, $P=P(T,V)$.
The research demonstrates that both gas models exhibit distinct behavioral patterns at different cosmic scales. At minimal volumes, they manifest as pressure-less systems, while at expanded volumes, they maintain a consistent negative pressure. This behavior effectively illustrates the cosmic transition from a matter-dominated era to one governed by cosmological constant dynamics. Our thermodynamic framework establishes specific constraints within which the thermal equation of state operates, particularly noting its temperature dependence. The thermal range is bounded by $0 < T < T^*$, where $T^*$ represents the maximum attainable temperature for both gas models at minimal volumes. The investigation of cosmic acceleration reveals significant insights through various cosmological parameters. The deceleration parameter $q$ exhibits positive values at small volumes but transitions to negative values as volume increases, confirming a shift from decelerated to accelerated expansion in the late universe. The dark energy equation of state parameter $\omega$ maintains negative values with an increasing trend during cosmic expansion for both models. The density parameter $\Omega$ corroborates this evolutionary pattern, indicating a transition from matter to dark energy dominance near $z \approx 0.5$.
\\Our thermal analysis, conducted using normalized parameters, reveals several key features:
\begin{itemize}
\item The thermal equation demonstrates pure temperature dependence
\item Both gas models maintain thermodynamic stability during expansion
\item The expansion process proceeds without critical points or phase transitions
\item Pressure decreases consistently with universal expansion
\item The specific heat capacity at constant volume ($C_V$) remains positive
\item The speed of sound maintains positivity with $0<v_S^2<1$
\end{itemize}
These findings collectively establish that both the Modified Chaplygin-Jacobi and Modified Chaplygin-Abel gas models demonstrate thermodynamic stability while maintaining classical stability against perturbations, providing a robust framework for understanding cosmic evolution. We have also performed observational data analysis using CC+BAO and Pantheon+SH0ES datasets. The information criterion shown in Tab.~\ref{tab:MCJG_MCAG_model_metrics} indicates that both the model are observationally viable. In summary, both the MCJG and MCAG models exhibit similar trends in parameter values between the Pantheon+SH0ES and CC+BAO datasets, but with notable differences in the magnitude of certain parameters, particularly $\Omega_{m0}$, $\omega_m$, $\alpha$, and $V_0$. These variations reflect differing assumptions about the roles of matter, dark energy, and potential interactions in the two models.

\section*{Acknowledgments}
D.M. is thankful to CSIR, Govt. of India for providing Senior Research Fellowship (No. 08/003(0145)/2021-EMR-I).

\section*{Data Availability Statement}
The data generated or analysed in this study are contained within the published article. No new data were created or examined during this research.

\section*{ Conflict of Interest Statement}
The authors declare that this research was carried out without any commercial or financial relationships that might be perceived as a potential conflict of interest.

\section*{Author Contributions}
Conceptualization, B.C. and T.M.; Methodology, T.M. and B.C.; Software, T.M. and B.C.; Formal Analysis, B.C., T.M., and D.M.; Writing—Original Draft Preparation, B.C. and T.M.; Writing—Review and Editing, D.M. and U.D.; Visualization, T.M. and B.C.; Supervision, U.D. All authors have read and agreed to the published version of the manuscript.

\bibliography{references}       
\bibliographystyle{elsarticle-num}  

\end{document}